\documentclass[iop]{emulateapj}
\usepackage{amsmath, ulem,longtable}
\usepackage[breaklinks,colorlinks,urlcolor=blue,citecolor=blue,linkcolor=blue]{hyperref}
\begin{document}

\title{SDSS-IV eBOSS Spectroscopy of X-ray and WISE AGN in Stripe 82X: Overview of the Demographics of X-ray and Mid-Infrared Selected Active Galactic Nuclei}

\author{Stephanie M. LaMassa$^1$, Antonis Georgakakis$^2$, M. Vivek$^{3,4}$, Mara Salvato$^5$, Tonima Tasnim Ananna$^{6,7}$, C. Meg Urry$^{6,7}$, Chelsea MacLeod$^8$, Nicholas Ross$^9$}
\affil{$^1$Space Telescope Science Institute, 3700 San Martin Drive, Baltimore, MD 21210;
$^2$National Observatory of Athens, Vas. Pavlou \& I. Metaxa, 15236 Penteli, Greece;
$^3$Department of Astronomy and Astrophysics, The Pennsylvania State University, 525 Davey Lab, University Park, PA 16802, USA;
$^4$Department of Physics \& Astronomy, University of Utah, Salt Lake City, UT 84112, USA;
$^5$Max-Planck-Institut f\:ur Extraterrestrische Physik, Garching, Germany;
$^6$Department of Physics, Yale University, P.O. Box 201820, New Haven, CT 06520-8120, USA;
$^7$Yale Center for Astronomy and Astrophysics, P.O. Box 208121, New Haven, CT 06520, USA;
$^8$Harvard-Smithsonian Center for Astrophysics, 60 Garden Street, Cambridge, MA, 02183, USA; 
$^9$Institute for Astronomy, University of Edinburgh, Royal Observatory, Blackford Hill, Edinburgh EH9 3HJ, UK 0000-0003-1830-6473}

\begin{abstract}
  We report the results of a Sloan Digital Sky Survey-IV eBOSS program to target X-ray sources and mid-infrared-selected {\it WISE} AGN candidates in a 36.8 deg$^2$ region of Stripe 82. About half this survey (15.6 deg$^2$) covers the largest contiguous portion of the Stripe 82 X-ray survey. This program represents the largest spectroscopic survey of AGN candidates selected solely by their {\it WISE} colors. We combine this sample with X-ray and {\it WISE} AGN in the field identified via other sources of spectroscopy, producing a catalog of 4847 sources that is 82\% complete to $r\sim22$. Based on X-ray luminosities or {\it WISE} colors, 4730 of these sources are AGN, with a median sample redshift of $z\sim1$. About 30\% of the AGN are optically obscured (i.e., lack broad lines in their optical spectra). BPT analysis, however, indicates that 50\% of the {\it WISE} AGN at $z<0.5$ have emission line ratios consistent with star-forming galaxies, so whether they are buried AGN or star-forming galaxy contaminants is currently unclear. We find that 61\% of X-ray AGN are not selected as MIR AGN, with 22\% of X-ray AGN  undetected by {\it WISE}.  Most of these latter AGN have high X-ray luminosities ($L_{\rm x} > 10^{44}$ erg s$^{-1}$), indicating that MIR selection misses a sizable fraction of the highest luminosity AGN, as well as lower luminosity sources where AGN heated dust is not dominating the MIR emission. Conversely, $\sim$58\% of {\it WISE} AGN are undetected by X-rays, though we do not find that they are preferentially redder than the X-ray detected {\it WISE} AGN.
\end{abstract}


\section{Introduction}
Active galactic nuclei (AGNs) serve as signposts of accreting supermassive black holes (SMBHs) across the Universe. Multi-wavelength selection of AGN is necessary for a complete picture of SMBH growth and evolution, mitigating selection biases that are inherent in any one band. Optical AGN selection favors Type 1 AGN, or those where we have a direct view of the accretion disk and associated broad line region. The typical blue colors of these Type 1 AGN and point-like morphology serve as a basis for ground-based targeting from optical spectroscopic surveys, such as the Sloan Digital Sky Survey \citep[SDSS,][]{york}, garnering hundreds of thousands of confirmed AGN \citep[e.g.,][]{paris}. However, optical surveys are biased against obscured AGN, where the accretion disk and broad line region are hidden behind large amounts of dust and gas on circumnuclear to galactic scales. Though these Type 2 AGN can be identified in nearby (i.e., $z<0.5$) galaxies on the basis of the ratios of their narrow emission lines using the so-called ``BPT'' diagram \citep{bpt}, they are more challenging to efficiently detect at larger distances, requiring infrared spectroscopic follow-up to observe the traditional BPT emission lines \citep{kewley2013a,kewley2013b}, or alternate diagnostics \citep{trouille,juneau,lamareille}.

X-rays, produced in a hot corona around the accretion disk, provide a direct probe of SMBH fueling. This energetic emission pierces through optically obscuring dust, but becomes attenuated at high gas column densities, especially at Compton thick levels ($N_H > 1.25 \times 10^{24}$ cm$^{-2}$) where they will appear X-ray weak \citep{bassani,heckman,lamassa2009,lamassa2011}, even at the highest X-ray energies \citep{lansbury2014,lansbury2015}.

AGN heated dust emits at mid-infrared energies, imparting a characteristic powerlaw shape to the spectral energy distribution, dominating over host galaxy star formation in powerful AGN \citep{lacy,stern2005,donley2012}. Mid-infrared (MIR) color selection then becomes a powerful tool to recover obscured AGN missed by optical and X-ray selection, though contamination from star-forming galaxies can be considerable at fainter fluxes \citep{barmby,cardamone,mendez}. Additionally, AGN at fainter luminosities are missed by mid-infrared selection \citep{mateos,menzel,lamassa2016b}.

A combination of AGN samples selected via independent methods is then crucial to understand selection effects and provide a comprehensive view of cosmic black hole growth. While survey fields, like GOODS \citep{alexander,comastri,xue,luo} and COSMOS \citep{scoville,hasinger,elvis,civano} have a wealth of multi-wavelength data where such analysis can be done, these cover a relatively small volume of the Universe with survey areas of $\sim$0.13 deg$^2$ and $\sim$2.2 deg$^2$, respectively. To find a representative sampling of rare AGN, wide-area surveys, which cover a large volume of the Universe, are necessary, complementing the AGN population found in smaller area fields.

Stripe 82X is such a wide-area X-ray survey, covering $\sim$31 deg$^2$ of the legacy SDSS Stripe 82 field \citep{lamassa2013a,lamassa2013b,lamassa2016a,ananna}. Imaged $\sim$100 times as part of a supernova legacy program \citep{frieman}, the coadded depth in Stripe 82 is approximately 2 magnitudes deeper than any single SDSS scan \citep{annis,jiang,fliri}. Stripe 82 contains rich multiwavelength coverage, with ultraviolet data from {\it GALEX} \citep{morrissey}, near-infrared (NIR) data from UKIDSS \citep{hewett,casali,lawrence} and the Vista Hemisphere Survey \citep[VHS;][]{mcmahon}, MIR data from {\it Spitzer} IRAC \citep{timlin,papovich} and {\it WISE} \citep{wright}, far-infrared coverage from {\it Herschel} SPIRE \citep{viero}, and radio coverage at 1.4 GHz from FIRST \citep{becker95,helfand}. About half ($\sim$15.6 deg$^2$) of the Stripe 82X survey is contiguous, spanning from 14$^{\circ} <$ RA $< 28^{\circ}$ and -0.6$^{\circ} <$ Dec $<0.6^{\circ}$. This region was observed between 2014 and 2015 with {\it XMM-Newton} in response to a successful AO13 proposal \citep[PI: Urry;][]{lamassa2016a}, and reaches an 0.5-10 keV flux limit of $\sim10^{-14}$ erg s$^{-1}$ cm$^{-2}$ at half survey area.

This contiguous, homogeneously covered portion of the Stripe 82X survey provides an ideal dataset to assess whether AGN identified via various multi-wavelength selection methods represent unique populations, determine the types of AGN common across identification methods, and construct a bolometric quasar luminosity function to analyze how AGN evolve over cosmic time. Important first steps are to determine redshifts and classifications for the X-ray sources via spectroscopy and to create an independent sample of MIR-selected AGN in the same survey area. These samples can then be combined with optically-selected AGN from SDSS within this survey area for a multi-wavelength view of black hole growth.

A special eBOSS \citep{smee, dawson} program of SDSS-IV \citep{gunn,blanton} spectroscopically observed the {\it XMM-Newton} AO13 Stripe 82X field, targeting 849 SDSS counterparts to Stripe 82 X-ray sources from the catalog of \citet{lamassa2016a} and 1518 independently selected {\it WISE} AGN candidates \citep[based on their $W1-W2$, 3.4$\mu$m - 4.6$\mu$m, color;][]{assef2013} within the same survey area. In this catalog release paper, we describe the observations and success rate of the spectroscopic identifications. We combine this source list with other spectroscopically identified X-ray and {\it WISE} AGN in the field to create a nearly complete sample of X-ray and MIR-selected AGN to $r \sim 22$, and comment on the demographics of these populations. Throughout, we assume a cosmology of $H_{\rm 0}$ = 67.8 km s$^{-1}$ Mpc$^{-1}$, $\Omega_{m}$ = 0.37, $\Omega_{\Lambda}$ = 0.69 \citep{planck2015}.

\section{Target Selection}
The parent X-ray and {\it WISE} samples we used for the target selection are discussed in detail below. We identified SDSS counterparts to these sources using the maximum likelihood estimator \citep[MLE;][]{sutherland}, a statistical approach that accounts for the distance between an X-ray ({\it WISE}) source and potential multiwavelength counterparts within a pre-defined search radius, the magnitudes of the potential associations within that search radius, the magnitude distribution of background sources, and astrometric errors on the X-ray ({\it WISE}) coordinates and those of the potential multiwavelength counterparts. This algorithm computes a likelihood ratio ($LR$), which is the probability that the correct counterpart is found divided by the probability that an unassociated background source is there by chance. From $LR$, a reliability value ($R$) is then calculated for each source. This counterpart matching was done separately for the X-ray and {\it WISE} sources, where $R$ ($LR$) was used to distinguish between true counterparts and spurious associations for the X-ray ({\it WISE}) sources, as discussed below.

\begin{figure}[ht]
  \begin{center}
  \includegraphics[scale=0.44]{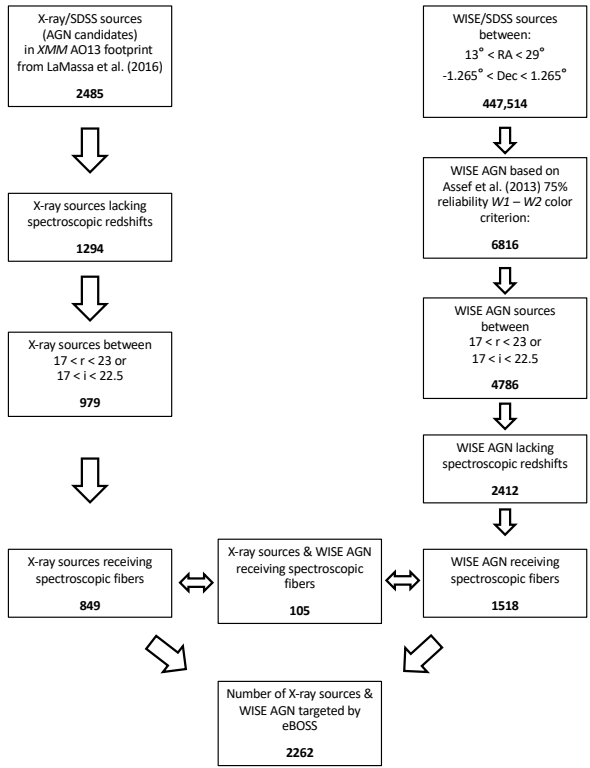}
  \caption{\label{targ_flowchart} Flowchart that illustrates how the X-ray and {\it WISE} AGN spectroscopic candidates were chosen. Note that the published catalog was slightly modified from the target list to use SDSS counterparts to X-ray sources from the \citet{ananna} Stripe 82X-multiwavelength catalog and to include {\it WISE} AGN candidates based on the updated color criteria of \citet{assef2018}.}
  \end{center}
\end{figure}

In total, 2262 SDSS counterparts to X-ray- and infrared-selected sources were targeted by the eBOSS Stripe 82X survey in the fall of 2015. Of these SDSS sources, 105 were both X-ray- and infrared- selected, 744 were detected only in X-rays, and 1413 were selected on the basis of their mid-infrared properties alone. Figure \ref{targ_flowchart} summarizes the selection criteria for both classes of objects, with further details in the following subsections.

After the eBOSS observations, an updated multi-wavelength matched catalog to the Stripe 82X survey was published in \citet{ananna}, using a deeper coadded SDSS catalog \citep{fliri} and matches to {\it Spitzer} data in the field \citep{timlin,papovich}. \citet{ananna} cross-matched these various multi-wavelength associations and reported the most likely counterpart to each X-ray source, resulting in $\sim$14\% discrepant associations compared with the \citet{lamassa2016a} catalog. Additionally, \citet{assef2018} published a {\it WISE} AGN catalog using slightly updated color selection criteria for the 90\% and 75\% reliability levels, leading to slight discrepancies between the eBOSS {\it WISE} AGN target list and the most up-to-date $W1-W2$ AGN definition.

When discussing the targeting procedure and inspecting the results of the pipeline, we preserve the original source lists, since success of the pipeline results depend on optical properties, regardless of why the source was included in the target list. However, we vet these lists to only retain the sources in the updated \citet{ananna} catalog and those that obey the revised \citet{assef2018} $W1-W2$ color criterion at the 75\% reliability level in the published catalog and when we comment on AGN demographics.

\subsection{X-ray}
\begin{figure*}[ht]
  \begin{center}
  \includegraphics[scale=0.6]{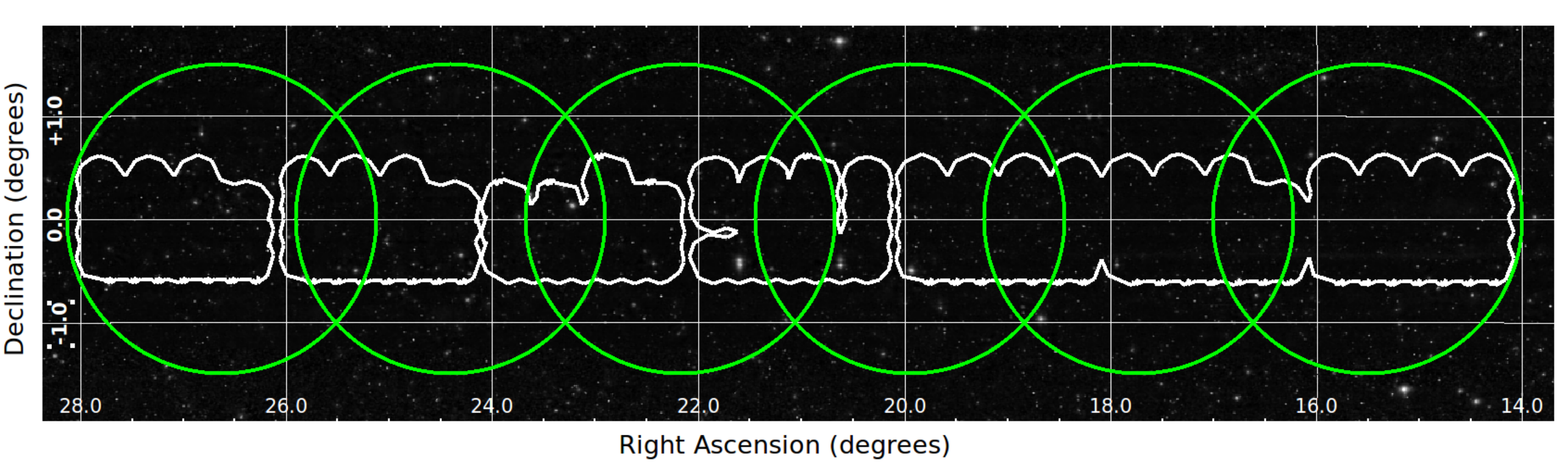}
  \caption{\label{fig:survey_layout} Layout of the field observed by the Stripe 82X eBOSS program. The green circles overlaid on the SDSS image mark the positions of the six SDSS-IV  spectroscopic plates of this program. The size of each circle is 3$^{\circ}$ in diameter, corresponding to the field of view of the SDSS spectroscopic plates which covers 36.8 deg$^2$ of Stripe 82. The white polygon demarcates the region covered by the 15.6 deg$^2$ AO13 {\it XMM-Newton} observations of the Stripe 82X survey \citep{lamassa2016a}.}
  \end{center}
\end{figure*}

The X-ray sample is culled from the 15.6 deg$^2$ portion of the 31.3 deg$^2$ Stripe 82 X-ray survey that was observed with {\it XMM-Newton} in AO13 \citep{lamassa2016a}. The full X-ray coverage in Stripe 82X includes $\sim$4.6 deg$^2$ of {\it XMM-Newton} observations from AO10 and archival {\it XMM-Newton} and {\it Chandra} observations in the field \citep{lamassa2013a,lamassa2013b} that were not observed with this SDSS-IV eBOSS spectroscopic program since the fields were mostly not contiguous with the {\it XMM-Newton} AO13 footprint. The X-ray-selected sources are from the {\it XMM-Newton} AO13 program as well as two archival {\it Chandra} observations and one archival {\it XMM-Newton} observation that overlapped the footprint of the SDSS spectroscopic plates. The coverage of the eBOSS ``Stripe 82X'' survey region is somewhat larger than the 15.6 deg$^2$ {\it XMM-Newton} AO13 survey area (see Figure \ref{fig:survey_layout}).

The details of the MLE matching are discussed in \citet{lamassa2013a,lamassa2016a}, but in brief, SDSS associations were identified within a 7$^{\prime\prime}$ search radius of an {\it XMM-Newton} source \citep{brusa2010}, or within a 5$^{\prime\prime}$ of a {\it Chandra} source \citep{civano2012}. The X-ray sources were originally matched to both the SDSS single-epoch imaging catalog and the coadded SDSS catalogs of \citep{jiang}, which reach a depth of  $r\approx 24.6$\,mag (AB), i.e. about two magnitudes deeper than the SDSS single-epoch data: if a source was found in the SDSS single-epoch imaging, we retained that match to enable efficient querying of the web-based SDSS database, otherwise we reported the magnitude(s) from the coadded catalog (if a reliable counterpart was found in this deeper catalog). We empirically determined a reliability threshold above which we accepted an SDSS source as a counterpart to an X-ray source: we shifted the X-ray positions by random amounts, repeated the MLE counterpart matching, and defined a critical reliability cutoff where the spurious association fraction (i.e., matches to randomized positions) was $\sim$10\% of the matches to actual X-ray sources. Any SDSS sources with a reliability value above this threshold was considered an X-ray counterpart.

\begin{figure}[ht]
  \begin{center}
  \includegraphics[scale=0.55]{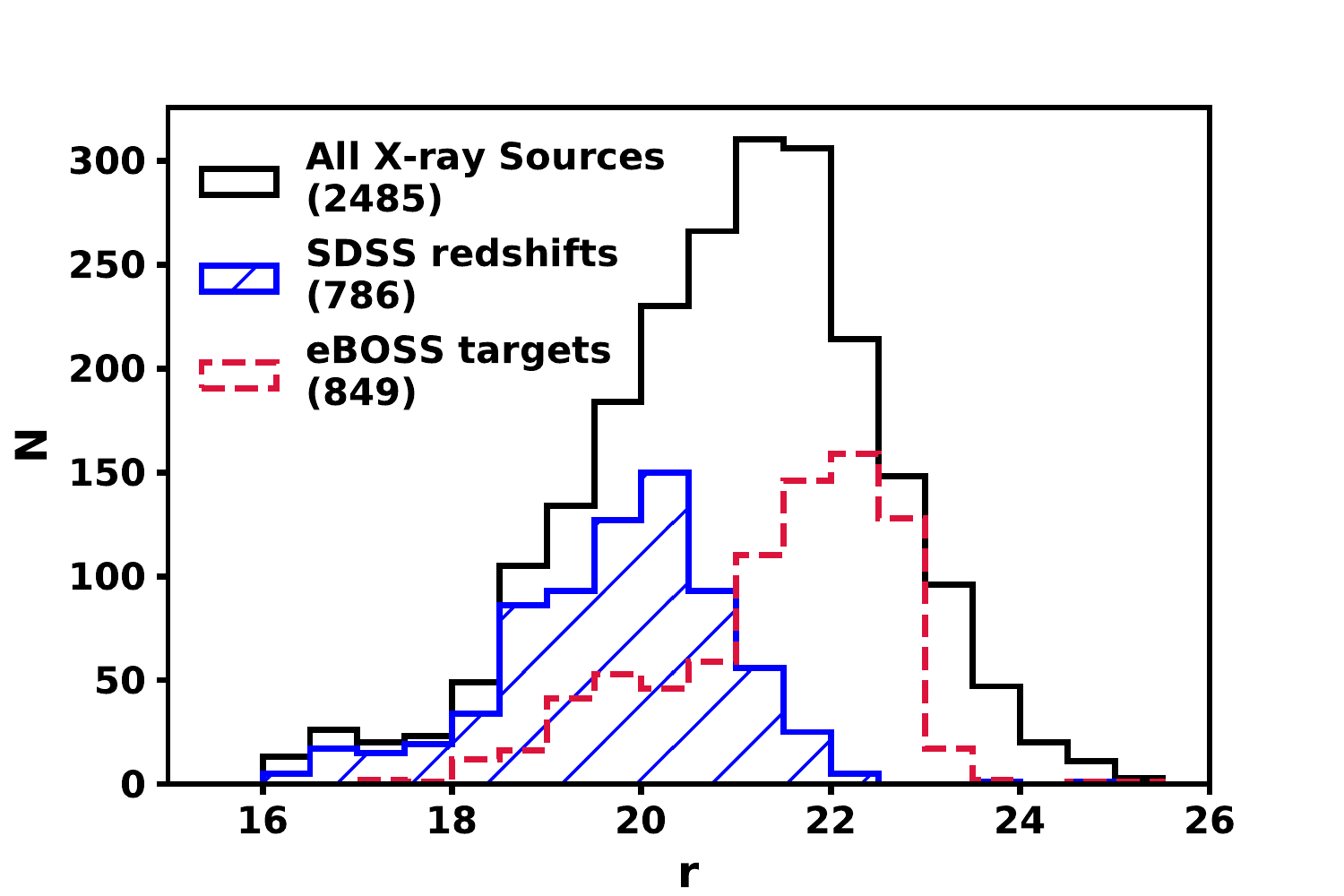}
  \caption{\label{rhist_xray} SDSS $r$-band optical magnitude distribution of the X-ray sources within the Stripe 82X {\it XMM-Newton} AO13 footprint (solid black histogram), with the subset of sources having pre-existing public spectroscopic redshifts from SDSS (blue hatched histogram) and targeted by this eBOSS program (red dashed histogram) highlighted; an additional 405 SDSS/X-ray sources have redshifts from other surveys and SDSS data that were proprietary at the time of the observations. The eBOSS targets are generally fainter than those with existing spectroscopy. }
  \end{center}
\end{figure}

Within the SDSS-IV eBOSS Stripe 82X survey area, there are 2485 SDSS sources that are identified as counterparts to X-ray sources in \citet{lamassa2016a}. Of these, 1191 had pre-existing spectroscopic redshifts, with 786 from previous SDSS programs  \citep[e.g., SDSS Data Releases 8 - 13 and previous releases of the SDSS quasar catalog;][]{albareti,alam2015,sdssdr8,sdssdr7} and the remainder from 2SLAQ \citep{croom}, 6dF \citep{jones2004, jones2009}, and proprietary SDSS redshifts at the time the target list was generated (which are now publicly available in SDSS Data Release 14 \citep{sdssdr14}). Of the 1294 X-ray/SDSS sources lacking redshifts, we imposed the following magnitude cuts to maximize the success rate of the SDSS-IV eBOSS program: $17 < r < 23$ or $17 < i < 22.5$, leaving us with 979 sources in the parent target list. Of these sources, 849 received fibers during the tiling process. Figure \ref{rhist_xray} shows the $r$-band magnitude distribution of the sample of the SDSS counterparts to X-ray sources within the SDSS-IV eBOSS Stripe 82X survey footprint, highlighting the sources with pre-existing SDSS spectroscopy that were public at the time of the observations and those targeted by the eBOSS survey.

\subsection{Mid-Infrared}

The starting point for the mid-infrared selection of AGN was the ALLWISE data release catalog (hereafter {\it WISE}), which combines data from the {\it WISE} cryogenic and NEOWISE missions \citep{Mainzer2011} as well as the post-cryogenic survey phases. Since the circular SDSS spectroscopic plates have a diameter of 3$^{\circ}$, six plates were needed to cover the width of the {\it XMM-Newton} AO13 Stripe 82X strip (see Figure \ref{fig:survey_layout}). The SDSS plates cover a greater area than the {\it XMM-Newton} AO13 footprint (36.8 deg$^2$), and the exact tiling strategy was only finalized after the parent target lists were generated. Hence, the {\it WISE} target list was chosen to cover the general area of the SDSS plates, with the exact targets chosen in the tiling process.

To create this master list, we selected {\it WISE} sources with right ascensions between 13$^{\circ} \leq$ RA $\leq$ 29$^{\circ}$ and declinations between -1.265$^{\circ} \leq$ Dec $\leq$ 1.265$^{\circ}$, which corresponds to the approximate width of Stripe 82 where deep coadded optical photometry is available. We further excluded potentially spurious {\it WISE} sources as well as sources with MIR photometry and/or positions affected by image artifacts by requiring that the contamination and confusion flag ({\sc cc\_flags}) of the {\it WISE} catalog equals zero in all four WISE spectral bands. 

These selections yield a total of 543,584 {\it WISE} sources. These are matched to optical counterparts using the Stripe 82 co-added catalogs presented by \cite{jiang}, using the MLE methodology as implemented in \citet{brusa2007}. The maximum radius within which potential counterparts are search for is set to 2$^{\prime\prime}$.  This limit is motivated by the sub-arcsec positional accuracies of both the {\it WISE} and \citet{jiang} catalogs. A likelihood ratio cut of  $LR>0.2$ is adopted for the WISE optical counterparts, which yields identifications for 82\% of the {\it WISE} sample (i.e., 447,514 sources) with an expected spurious fraction of $<3$\%. 

\begin{figure}[ht]
  \begin{center}
  \includegraphics[scale=0.55]{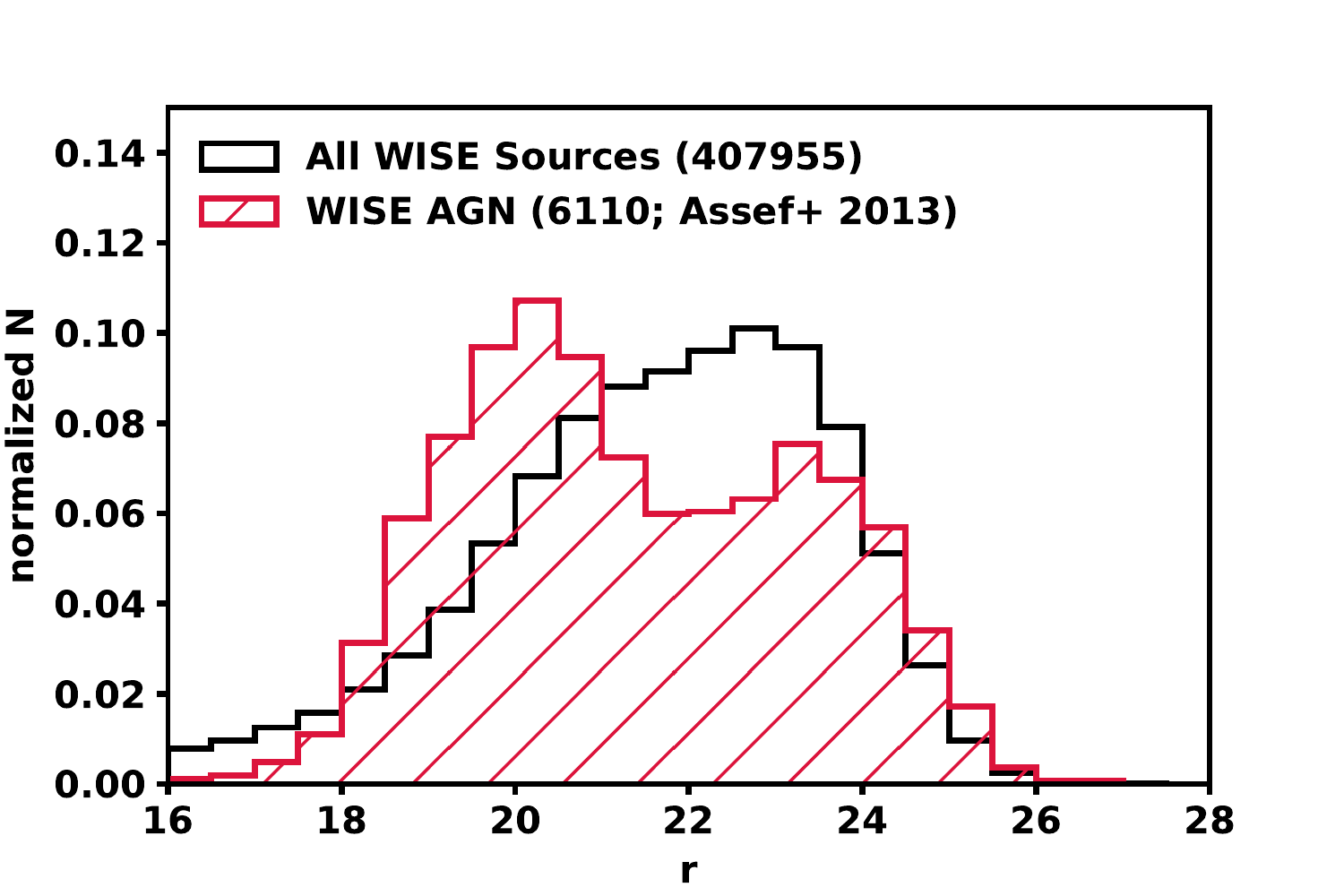}
  \caption{\label{fig:dndm_allwise} $r$-band optical magnitude distribution of the SDSS counterparts to the {\it WISE} sources in the Stripe 82 subregion that brackets the approximate area covered by this SDSS-IV eBOSS Stripe 82X program (13$^{\circ} \leq$ RA $\leq$ 29$^{\circ}$, -1.265$^{\circ} \leq$ Dec $\leq$ 1.265$^{\circ}$). The optical photometry is from \protect\cite{jiang}. The black histogram is for all {\it WISE} sources in the area while the red hatched histogram shows the magnitude distribution of the {\it WISE} AGN subsample selected using the $W1-W2$ color selection for 75\% reliability from \protect\cite{assef2013}; there are additional {\it WISE} sources and {\it WISE} AGN not shown here since they were not detected in the $r$-band in the \citet{jiang} catalog. }
  \end{center}
\end{figure}

\begin{figure}[ht]
  \begin{center}
  \includegraphics[scale = 0.6]{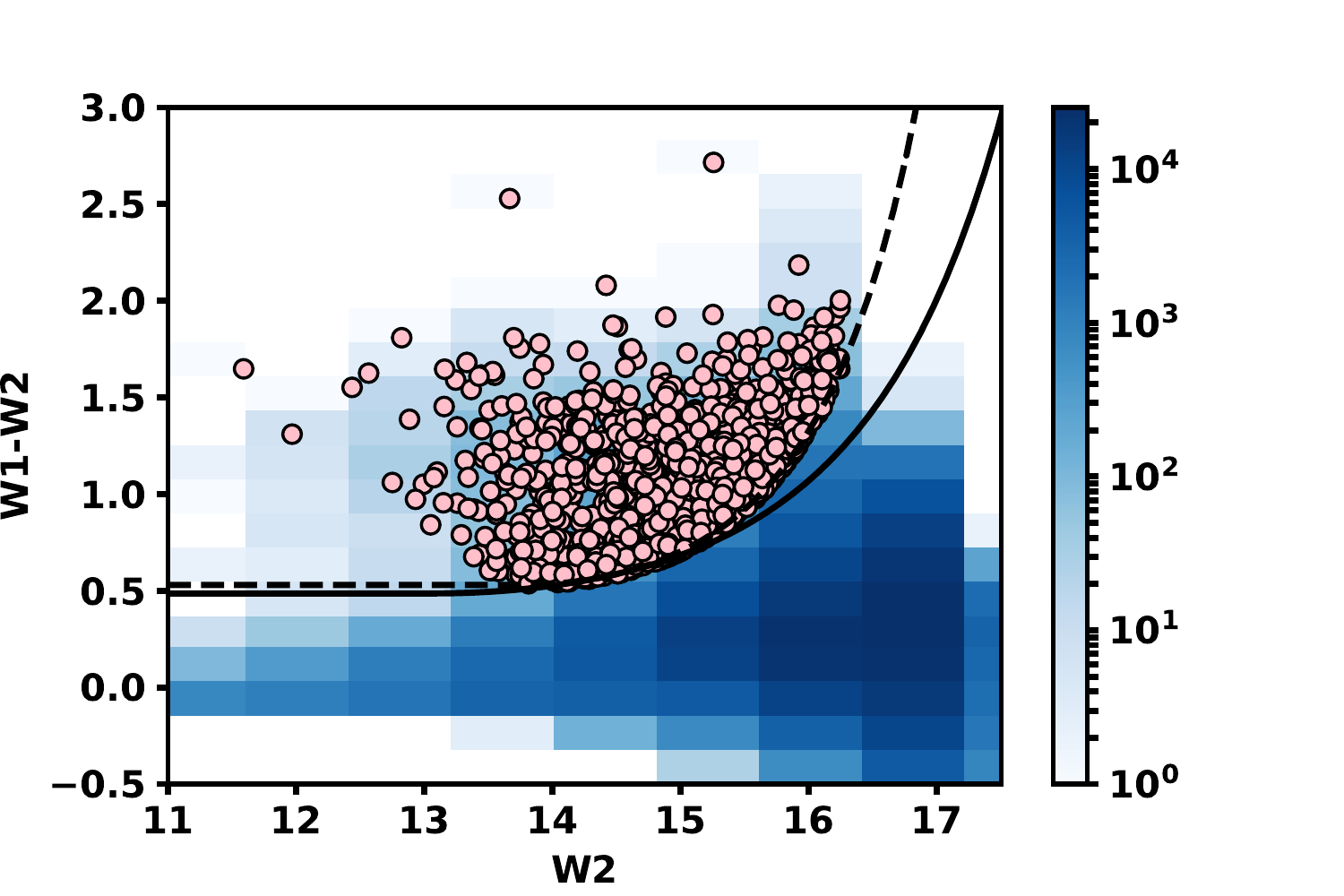}
  \caption{\label{w1w2_select} {\it WISE} $W1-W2$ color versus $W2$ magnitude for {\it WISE} sources within the SDSS-IV eBOSS Stripe 82X target area that obey the optical magnitude cuts in the main text ($r < 23$ or $i < 22.5$). The dashed line indicates the \citet{assef2013} $W1-W2$ color cut for 75\% reliability, which was used to make the target list; the solid line shows the revised \citet{assef2018} criterion, which we use to generate the final catalog and to comment on source demographics. The targets are shown by the pink circles overplotted on the total number of {\it WISE} sources, as indicated by the color bar.}
  \end{center}
\end{figure}

Figure \ref{fig:dndm_allwise} presents the $r$- magnitude distribution of the {\it WISE} optical counterparts; 39,559 SDSS counterparts were not detected in the $r$-band and are thus not included in this plot. Also shown in this figure is the distribution for 6110 {\it WISE} AGN candidates with $r$-band detections and {\it WISE} $W1 - W2$ colors redder than the  75\% reliability color-cut defined by \cite{assef2013}, as shown in Figure \ref{w1w2_select}; an additional 706 {\it WISE} AGN are not shown since they lack $r$-band detections. These sources have a bi-modal $r$-band magnitude distribution. The optically-faint peak of the distribution in Figure \ref{fig:dndm_allwise} may include a large fraction of obscured AGN \citep[e.g.][]{DiPompeo2014}, where the optical range of the Spectral Energy Distribution is dominated by host galaxy rather than AGN light. 

\begin{figure}[ht]
  \begin{center}
  \includegraphics[scale=0.55]{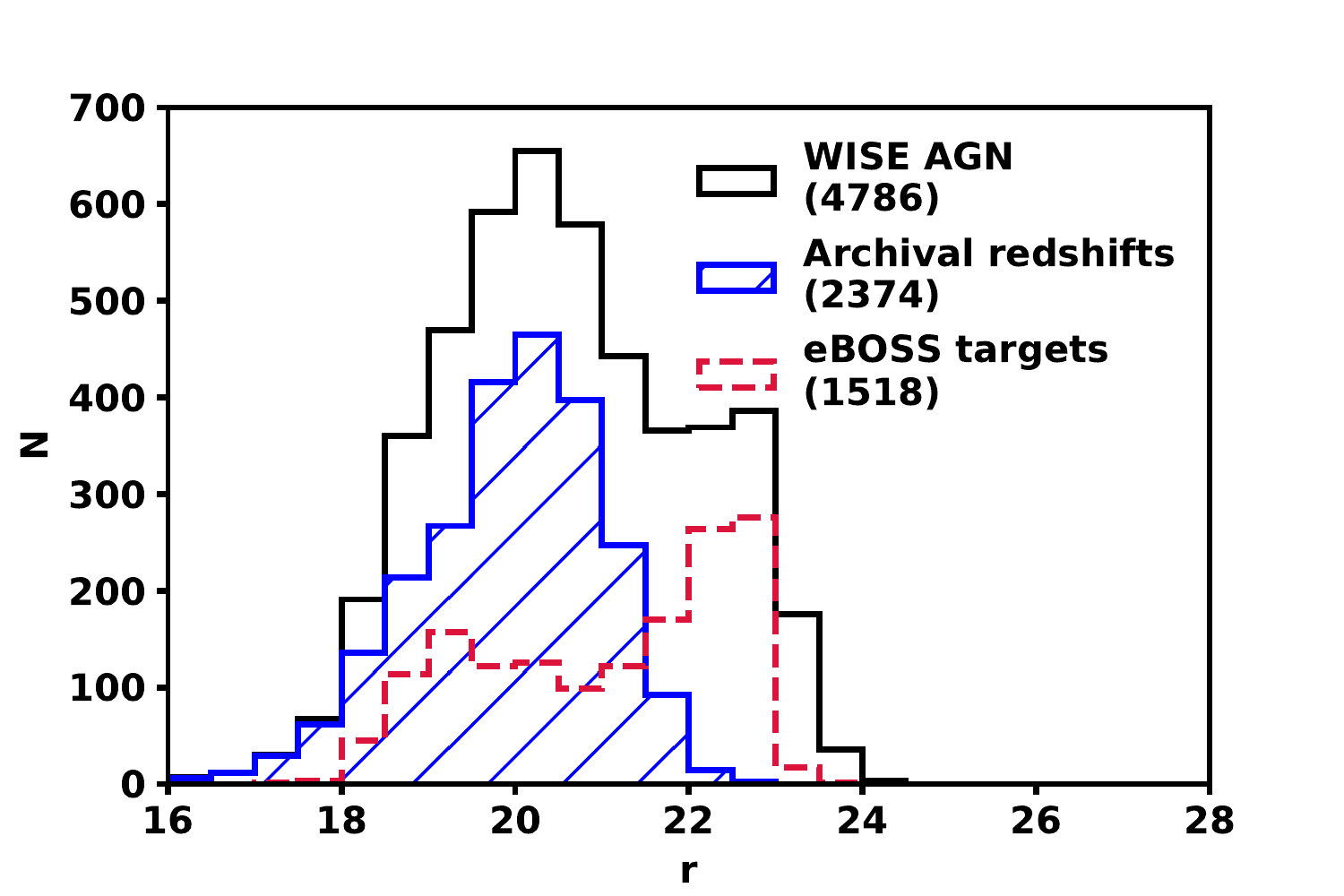}
  \caption{\label{fig:dndm_wiseagn} $r$-band optical magnitude distribution of the {\it WISE} AGN that are selected using the 75\% reliability criterion of \protect\cite{assef2013}, spatially overlap the targeting region of the SDSS-IV eBOSS Stripe 82X program, and are brighter than $r < 23$ or $i < 22.5$ (black solid histogram). The blue hatched histogram corresponds to sources with spectroscopy from SDSS programs prior to this eBOSS survey, and the red-dashed region shows the magnitude distribution of the {\it WISE} AGN candidates that received fibers in the SDSS tiling process and were thus targeted by the eBOSS program.}
  \end{center}
\end{figure}

Targets for follow-up spectroscopy were selected to lie within the 75\% reliability {\it WISE} $W1 - W2$ color-wedge defined by \cite[][; see Figure \ref{w1w2_select}]{assef2013} and have optical counterparts with magnitudes brighter than $r=23$, or $i=22.5$\,mag, in the \cite{jiang} co-added catalog, leaving 4786 sources. These optical limits are a trade-off between depth, to explore the nature of  the optically-faint WISE AGN candidates (e.g. Figure \ref{fig:dndm_allwise}), and sufficient signal-to-noise of the resulting SDSS spectra to measure reliable redshifts, at least in the case of emission line galaxies and/or AGN \citep[][]{menzel, Raichoor2017, Delubac2017}. Spectroscopy is available for 2374 of the {\it WISE} AGN from previous SDSS programs. These sources were not targeted as {\it WISE}-selected AGN in the eBOSS survey unless they qualified for repeat observations to explore QSO optical spectral variability (see Section \ref{add_targs}). The remaining 2412 sources were potential spectroscopic targets as {\it WISE} AGN candidates; 1518 received fibers in the tiling process. The optical magnitude distribution of these sources is shown in Figure \ref{fig:dndm_wiseagn}. 

\subsection{Additional Targets}\label{add_targs}
Additional sources were added to the target list to make use of all available spectroscopic fibers during the tiling process (see Table \ref{tiling_info} for a summary). These ``filler" targets included:

\begin{itemize}
  
\item quasar candidates from the photometric redshift catalogs of:
  
  \begin{itemize}
  \item \citet{richards}, using SDSS and {\it WISE} photometry (``S82X\_RICHARDS15\_PHOTOQSO\_TARGET"),
    
  \item \citet{peters}, using optical photometry and variability in Stripe 82 (``S82X\_PETERS15\_COLORVAR\_TARGET");
  \end{itemize}
  
\item $z > 4$ quasar candidates from LSST \citep[][``S82X\_LSSTZ4\_TARGET"]{alsayyad};

\item changing-look AGN candidates, where the optical spectra may show disappearing or emerging broad Balmer lines between spectroscopic epochs \citep[e.g.,][]{denney,shappee,lamassa2015,ruan,runnoe,gezari,yang}, using the photometric variability cuts employed in \citet{macleod}, with or without an additional cut on the timing of the previous spectral epoch (``S82X\_CLAGN1\_TARGET'' or ``S82X\_CLAGN2\_TARGET'', respectively);

\item {\it WISE} AGN candidates from {\it WISE} forced photometry at the positions of known SDSS sources \citep{lang} that were otherwise not already in the {\it WISE} target list (``S82X\_UNWISE\_TARGET'');

\item  photometric variability selected quasar candidates from \citet[][``S82X\_SACLAY\_VAR'']{palanque-delabrouille};

\item quasar candidates selected on the basis of their SDSS and {\it Spitzer} colors, using a boosted decision tree algorithm (``S82X\_SACLAY\_BDT'');

\item  high redshift quasar candidates identified by defining drop-out regions in optical color and optical - {\it WISE} color parameter space \citep[``S82X\_SACLAY\_HIZ''; see][]{richards2002}.
\end{itemize}

The SDSS counterparts to the X-ray sources and the {\it WISE} AGN candidates are listed as ``S82X\_XMM\_TARGET'' and ``S82X\_WISE\_TARGET'', respectively, in Table \ref{tiling_info} and in the SDSS database. We focus on these targets exclusively when commenting on the results of the eBOSS program, and note that the spectra for the ancillary targets were made public in SDSS Data Release (DR) 14 \citep{sdssdr14}, with specific samples to be discussed in future papers (e.g., MacLeod et al., in prep.).

\subsection{Tiling}
The objective of the tiling process is to achieve a distribution of sources across a plate that maximizes the number of observed targets with a minimum number of plates \citep{dawson}. As the X-ray and {\it WISE} target density was fairly uniform across the region, we chose to tile six plates with fixed centers (see Figure \ref{fig:survey_layout}). These plates are identified with plate identification numbers ranging from 8788-8793.  Target selection algorithms resulted in an average number of 1304 targets per plate. Five percent of these target were eliminated either due to a possible knock out with an allocated high priority fiber or due to their high brightness.

 While tiling the fibers across the plates, we opted to treat each plate independently  so that there are repeat spectra of high-priority targets in the overlap regions. However, there are many targets that are relatively bright ($r < 22$), in which case the S/N in a nominal observation ($\sim$ 2 hrs) would be sufficient ("S82X\_BRIGHT\_TARGET").
Hence, we adopted a tiered-priority system for assigning fibers to the targets. The advantage of this system is that we could free up some fibers for additional targets by removing the bright objects from the overlap regions as each plate is successively tiled. For each plate, we carried out three rounds of tiling and in each round we assigned fibers to targets depending on the priority of the targets. 
Targets corresponding to ``S82X\_XMM\_TARGET", ``S82X\_WISE\_TARGET",  ``S82X\_LSSTZ4\_TARGET", ``S82X\_CLAGN1\_TARGET", and ``S82X\_CLAGN2\_TARGET" were  included in the  highest priority list. The next priority list includes targets corresponding to ``S82X\_SACLAY\_VAR\_TARGET", ``S82X\_SACLAY\_BDT\_TARGET", ``S82X\_RICHARDS15\_PHOTOQSO\_TARGET", and ``S82X\_PETERS15\_COLORVAR\_TARGET". The final priority list contains targets corresponding to ``S82X\_BRIGHT\_TARGET", ``S82X\_SACLAY\_HIZ\_TARGET", and ``S82X\_UNWISE\_TARGET".  Table ~\ref{tiling_info} lists  the number of available targets and the number of tiled targets for the different  target classes among the six plates. For each round, if all the higher priority targets were assigned fibers, the remaining fibers were allocated to the targets in the next priority. Three hundred fifty-three targets have additional spectra owing to the overlap of the plates. 

More specifically, and relevant to this catalog release, a total of 849 X-ray sources and 1518 {\it WISE} AGN candidates received spectroscopic fibers in the tiling process: 744 sources were X-ray only, 1413 sources were {\it WISE} only, and 105 sources were both X-ray and {\it WISE} targets, for a total of 2262 unique SDSS sources receiving fibers. Optically faint X-ray and {\it WISE} AGN candidates in overlapping regions between the neighboring plates were observed for twice the nominal exposure time, for 95 X-ray sources and 92 {\it WISE} sources.

\begin{deluxetable*}{lllllll}
  \tablewidth{0pt}
  \tablecaption{\label{tiling_info} SDSS Tiling Summary: Number and Fraction of Targets Receiving Spectroscopic Fibers on Each SDSS Plate\tablenotemark{1}}
\tablehead{ &  8788 & 8789 & 8790 & 8791 &	8792 & 8793          \\      
TARGET CLASS                       &      $N_{\rm tot}-N_{\rm tile}$	&	       $N_{\rm tot}-N_{\rm tile}$	     &     $N_{\rm tot}-N_{\rm tile}$	&      $N_{\rm tot}-N_{\rm tile}$	   &       $N_{\rm tot}-N_{\rm tile}$	&      $N_{\rm tot}-N_{\rm tile}$		 }
 \startdata 
                                          
S82X\_BRIGHT\_TARGET                 &	921 - 547	&	       832 - 520	     &     678 - 470 	&      593 - 451	   &       703 - 461 &      705 - 478 	 \\
S82X\_XMM\_TARGET               	   &     198 - 187    &	        197 - 178     &       142 - 132   &          75 -  71 &           169 - 162 &           208 - 195    \\
S82X\_WISE\_TARGET               	   &     324 - 303      &	        289 - 269     &       259 - 248   &         247 - 235 &           297 - 282 &           287 - 273     \\
S82X\_SACLAY\_VAR\_TARGET             &	273 - 157        &	       249 - 155      &      249 - 193    &        278 - 254 &          241 - 181  &          263 - 1829	\\
S82X\_SACLAY\_BDT\_TARGET             &	274 - 155        &	       241 - 166     &      228 - 175   &        205 - 183  &          197 - 146   &          207 - 133	\\
S82X\_SACLAY\_HIZ\_TARGET             &	154 -  18      &	       132 -  16      &      103 -  24    &         83 -  19  &          120 -  17 &          118 -  25 	\\
S82X\_RICHARDS15\_PHOTOQSO\_TARGET    &	 32 -  24        &	        40 -  30      &       30 -  23    &         29 -  29 &           42 -  40   &           42 -  35	\\
S82X\_PETERS15\_COLORVAR\_TARGET      &	225 - 132       &	       265 - 173     &      219 - 179     &        212 - 192  &          215 -  66  &          239 - 161	\\
S82X\_LSSTZ4\_TARGET                 &	 27 -  27         &	        18 -  18      &       23 -  23     &         27 -  27   &           19 -  19  &           25 -  25	\\
S82X\_UNWISE\_TARGET                 &	206 -  59        &	       170 -  49       &      125 -  37    &        127 -  48   &          173 -  54   &          152 -  62 	\\
S82X\_CLAGN1\_TARGET                 &	  2 -   2        &	         2 -   1      &        5 -   5     &          4 -   4 &            5 -   5  &            5 -   5 \\
S82X\_CLAGN2\_TARGET                 &	 23 -  21         &	        26 -  23      &       31 -  29     &         37 -  35  &           35 -  31  &           31 -  28 	\\
\hline \\
TOTAL                        	   &    1497 - 900       &	       1422 - 900     &      1228 - 900   &        1117 - 900  &          1282 - 900 &          1313 - 900  
\enddata
\tablenotetext{1}{$N_{\rm tot}$ is total number of targets overlapping the plate and $N_{\rm tile}$ is the number of targets that received spectroscopic fibers.}

\end{deluxetable*}

\section{Observations}
The six plates were observed to twice the depth as that of typical eBOSS observations from previous SDSS data releases \citep{dawson}. A plate is exposed for 15 minutes per exposure, with the number of exposures repeated until the square of the signal-to-noise, (S/N)$^2$, per pixel in all four cameras (two red and blue cameras for spectrograph one and spectrograph two) passed a pre-determined (S/N)$^2$ threshold. Standard eBOSS plates had a (S/N)$^2$ threshold of 10 and 22 in the blue and red cameras, respectively, for an object with $g$ = 22 and $i$ = 21, respectively \citep{dawson}; here the magnitudes are measured through the SDSS spectroscopic fibers.

As the Stripe 82X targets are relatively fainter compared to previous eBOSS targets, the Stripe 82X plates were exposed to a higher (S/N)$^2$ threshold of 20 and 44 for the red and blue cameras, respectively. While it required 19 exposures at 15 minutes an exposure for two plates (8790 and 8791) due to bad observing conditions, the remaining four plates had 8-12 exposures of 15 minutes apiece. All the observed data were run through the full SDSS pipeline, IDLSPEC2D v5\_10\_0. IDLSPEC2D extracts the spectrum corresponding to each exposure and combines the individual spectra to give a combined high SN spectrum for each target. While combining the individual spectra, the SDSS pipeline excludes those exposures which have (S/N)$^2$ less than 20\% of the (S/N)$^2$ of the best exposure. The pipeline rejected fiver exposures for plate 8791, but none for any of the remaining five plates.

\subsection{Pipeline Processing}

\noindent Both the {\it WISE} and X-ray selected targets of this eBOSS survey include a large fraction of optically faint sources (e.g. Figure \ref{fig:dndm_wiseagn}), close to the spectroscopic limit of 2-m class telescopes. Visual inspection was therefore deemed necessary to control the quality of the redshift measurements and optical spectral classifications for individual sources. 

The visual inspection of the WISE spectra proceeded in two stages. First a total of 5 human classifiers inspected the spectra observed on a single plate of the survey (plate number 8792, observed on Modified Julian Date, MJD, 57364). The classification included the determination of the source redshift, an assessment of the redshift-measurement reliability/quality and the assignment of a rough spectral class. The starting point of the visual inspection were the products of the SDSS spectral reduction pipeline version 5.10.0 \citep{Bolton2012}. These include among others a best-fit spectral template and the corresponding redshift for each source as well as a warning flag ({\sc zwarning}) raised in the case of bad, uncertain or more generically problematic redshift fits. The classifiers were presented with the best-fit pipeline products and had to decide whether they agreed or not, modify the redshift if they deemed necessary, assign a redshift quality flag ({\sc z\_conf}) and a spectral class ({\sc class\_person}). The possible values of the latter flags and their meaning are presented Table \ref{vi_scheme}. The visual inspection process enforced agreement of all classifiers on the redshift, quality and class of a given source. Discrepancies were discussed and settled in a reconciliation round, which result to a final list of redshifts for the sources targeted on plate number 8792 (Modified Julian Date 57364).

Based on these results, 2 of the classifiers (AG, SL) visually inspected the remaining WISE-selected AGN candidates (AG) and X-ray sources (SL) targeted by this eBOSS program; a third classifier (VM) reviewed uncertain redshifts for the X-ray sources and all three resolved any discrepant classifications via additional visual inspection and discussion. In the analysis that follows we use all redshifts with {\sc z\_conf}$\ge2$, i.e. spectra with at least a single identified feature.

\begin{deluxetable*}{lllll}
  \tablewidth{0pt}
  \tablecaption{\label{vi_scheme} Visual Inspection Classification Scheme}
  \tablehead{ &  & & \colhead{{\sc z\_conf}} & \\
 & \colhead{0} & \colhead{1} & \colhead{2} & \colhead{3} \\
    \colhead{{\sc class\_person}}}
 \startdata 
  
NONE            &  no vote/opinion    & Bad spectrum       & no signal (continuum or spectral features)         & continuum but no spectral features \\
QSO             &  \nodata              & \nodata                & single broad line  & $>1$ spectral features (including broad lines)  \\
Galaxy          &  \nodata               & \nodata               & single non-broad line  & $>1$ lines (non-broad)  \\
QSO BAL         &  \nodata               & \nodata                & BAL troughs only; uncertain $z$   & BAL troughs and (narrow) emission lines \\
Blazar         &  \nodata              & \nodata              & continuum, no features to measure $z$   & continuum \& spectral features to measure $z$   \\
Star           &  \nodata              & \nodata             & Stellar-like continuum  &  Stellar-like continuum \&  spectral features 
\enddata

\end{deluxetable*}

\section{Results}
Of the 2262 SDSS sources targeted, we verified or independently determined redshifts and classifications for 1769 objects (78\% success rate), with 1602 sources where {\sc z\_conf} = 3 and 167 sources where {\sc z\_conf} = 2. Of these sources, 591 are QSOs, 1129 are galaxies, and 49 are stars. We expect that the results from this eBOSS pilot program will inform observing strategies and efficient data quality control checks for future SDSS surveys and those from other ground-based observatories.

\begin{figure*}[ht]
  \begin{center}
  \includegraphics[scale=0.35,angle=270]{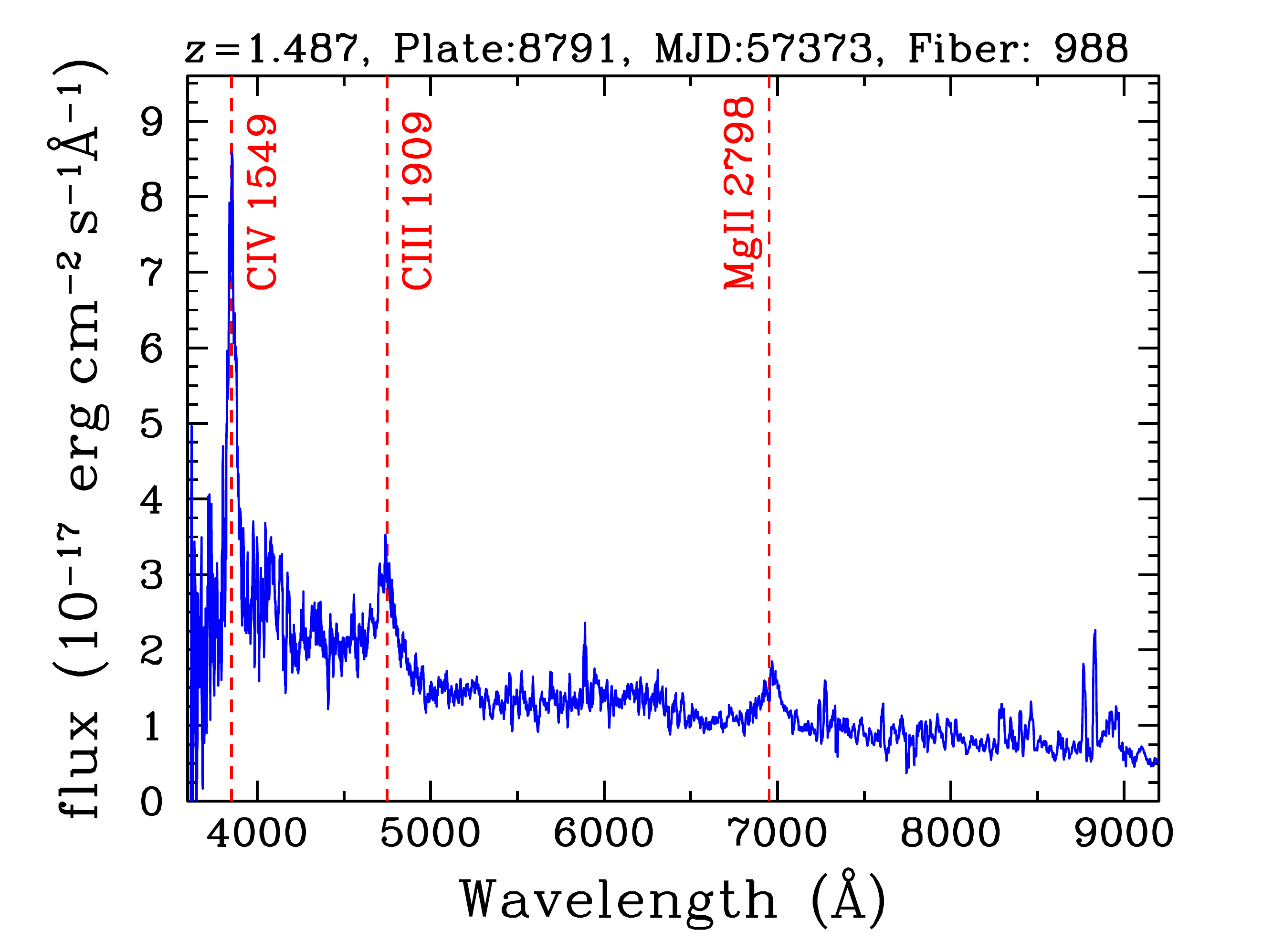}~
  \includegraphics[scale=0.35,angle=270]{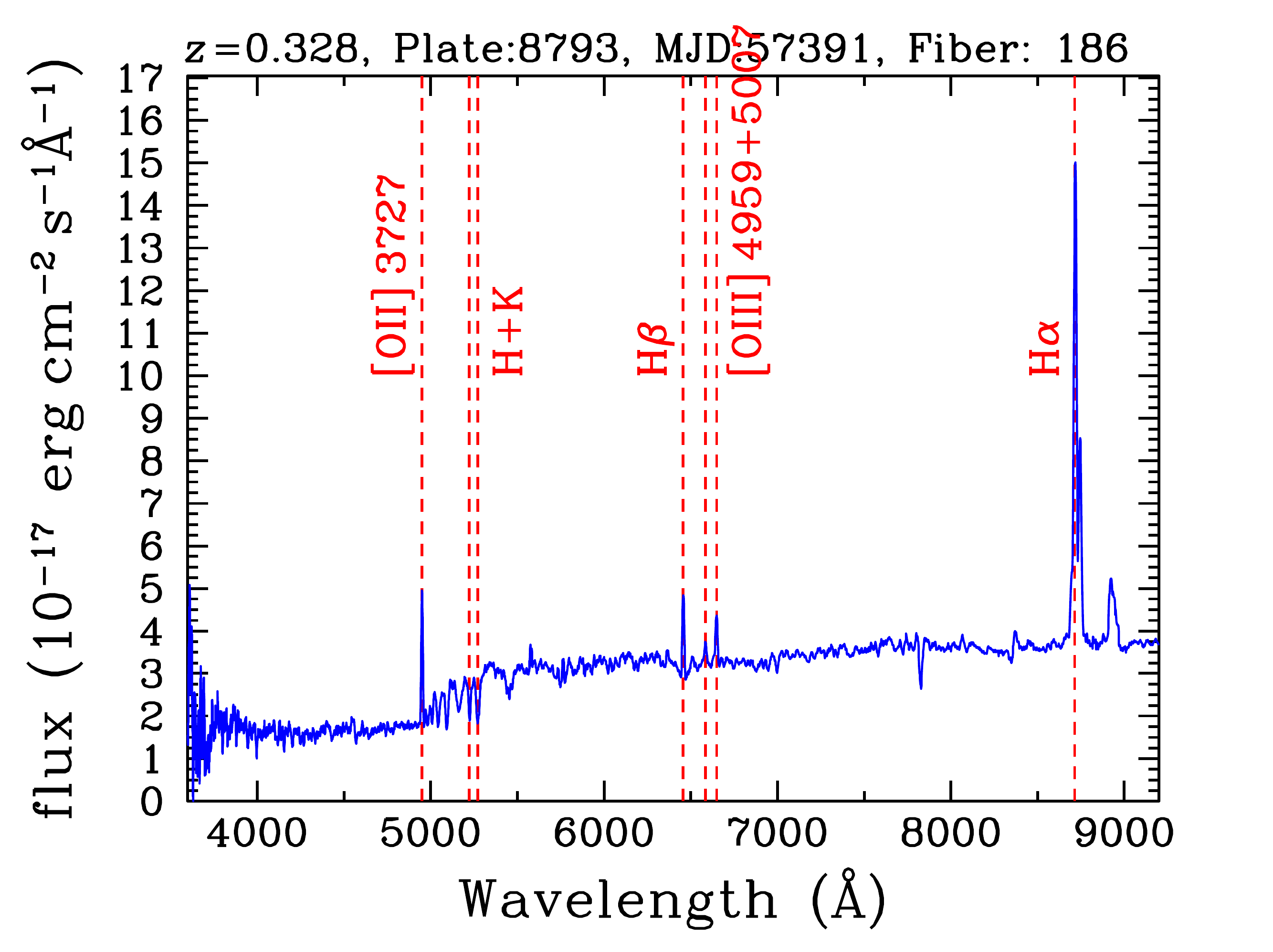}
  \includegraphics[scale=0.35,angle=270]{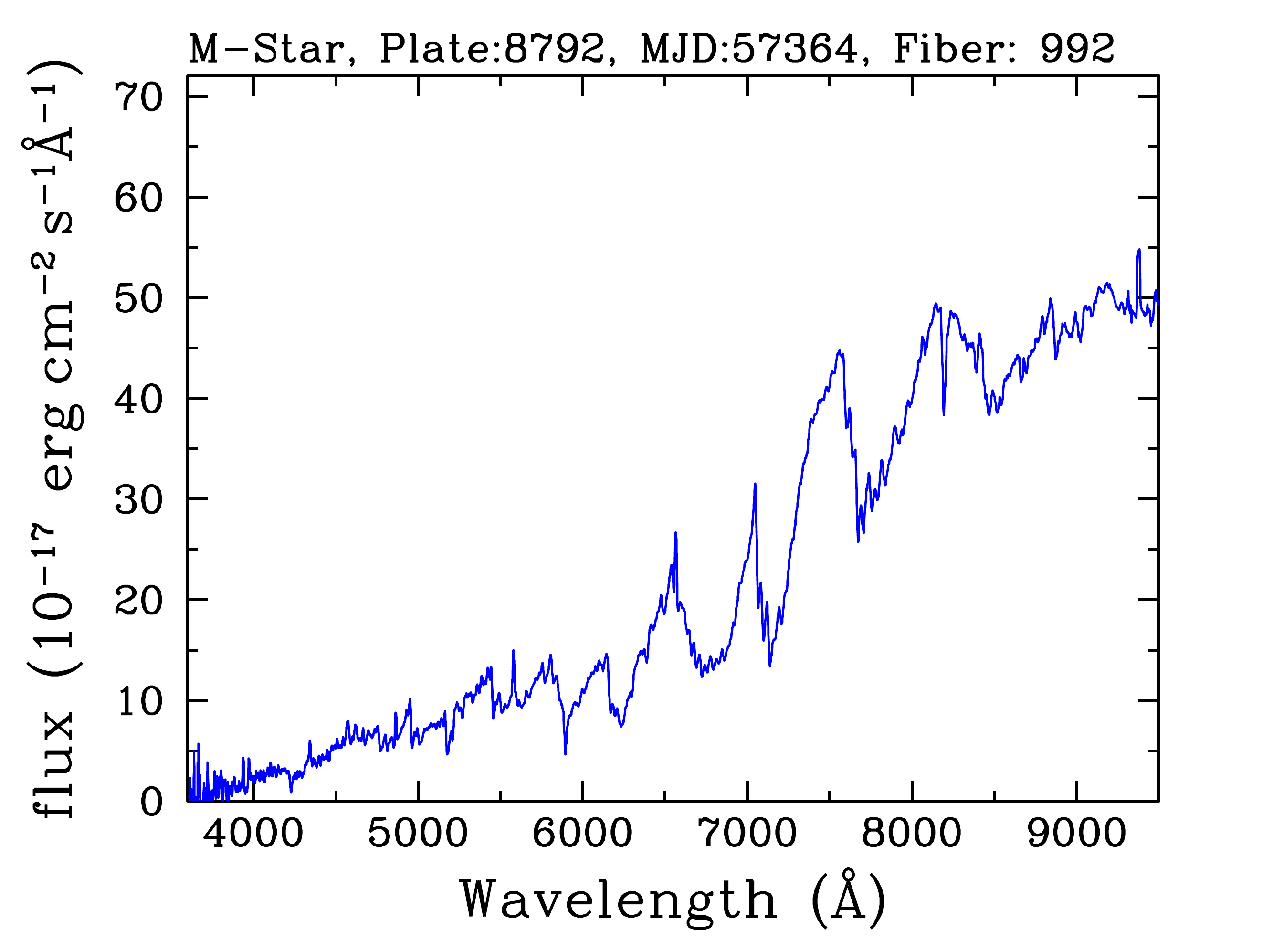}
  \caption{\label{sdss_spec} Example SDSS spectra of the various classes of objects detected in this eBOSS survey: ({\it top left}) Type 1 AGN at $z=1.487$ (MJD = 57373, fiber = 988, plate = 8791), ({\it top right}) SDSS galaxy at $z = 0.328$ (MJD = 57391, fiber = 186, plate = 8793), and ({\it bottom}) star (M-Dwarf; MJD = 57364, fiber = 992, plate = 8792); notable transitions in the spectra of the extragalactic sources are indicated. All three objects were detected by both {\it WISE} and in X-rays. Both the QSO and galaxy have X-ray luminosities consistent with AGN and {\it WISE} $W1-W2$ AGN colors at the 75\% reliability threshold \citep{assef2013}.}
  \end{center}
\end{figure*}

We show an example spectrum of each of these sources in Figure \ref{sdss_spec} to highlight the variety of objects detected by the eBOSS program; for reference, we include identifying information for these sources (MJD of observation, plate number, and fiber identification number) in the caption.

\subsection{Identification Success Rate}
\begin{figure}[ht]
  \begin{center}
  \includegraphics[scale=0.4]{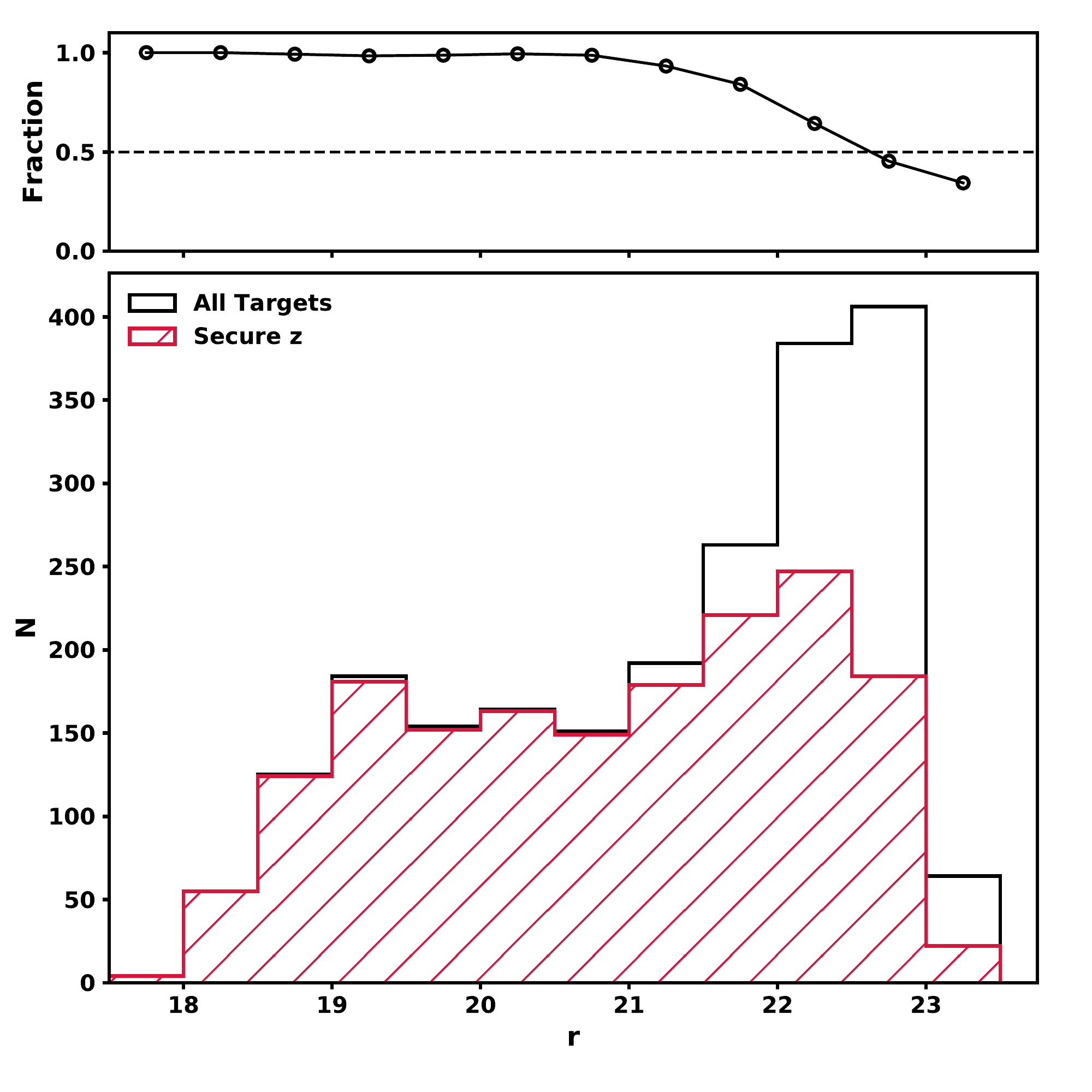}
  \caption{\label{success_rmag} SDSS $r$-band optical magnitude \citep{jiang} distribution of the X-ray sources and {\it WISE} AGN candidates targeted by the eBOSS program. The solid black line shows all targets while the red hatched histogram shows those sources who have reliable spectroscopic redshifts. The ratio of the two (i.e., the fraction of spectroscopic identifications as a function of $r$ magnitude) is shown in the top panel; a horizontal line at 50\% spectroscopic completeness is shown for reference. The fraction of identified sources drops below 50\% at $r > 22.5$.}
  \end{center}
\end{figure}

We explore the success rate of the pipeline as a function of optical magnitude in Figure \ref{success_rmag}, where we show the $r$-band magnitude distribution for all targets (solid black histogram) and the subset with reliable redshifts
(red hatched histogram; here we use the co-added optical photometry from \citet{jiang} for all sources). The top panel of Figure \ref{success_rmag} shows the fraction of sources identified as a function of their $r$ magnitude, with a horizontal line at 50\% completeness shown for reference.

We can identify more than half of the sample at $r < 22.5$. Even at the faintest magnitude limits (i.e., $22.5 < r < 23.5$), we are able to obtain reliable redshifts and classifications for above 37\% of the sample, which is a significant fraction. 

This success rate suggests that future SDSS programs can relax the nominal limiting magnitude constraint \citep[e.g., $r < 22$ in the eBOSS quasar survey;][]{myers} for deeper observations and targets expected to have emission lines, akin to the Stripe 82X eBOSS survey. Our results indicate that surveys from observatories that have larger aperture mirrors, like the 4-m Dark Energy Survey, will also be successful in obtaining spectroscopic redshifts for sources to faintness levels of $r \sim 24$ under similar observing conditions.

\subsection{Pipeline vs. Visual Inspection: Clues from Pipeline Flags}
\begin{figure}[ht]
  \begin{center}
  \includegraphics[scale=0.6]{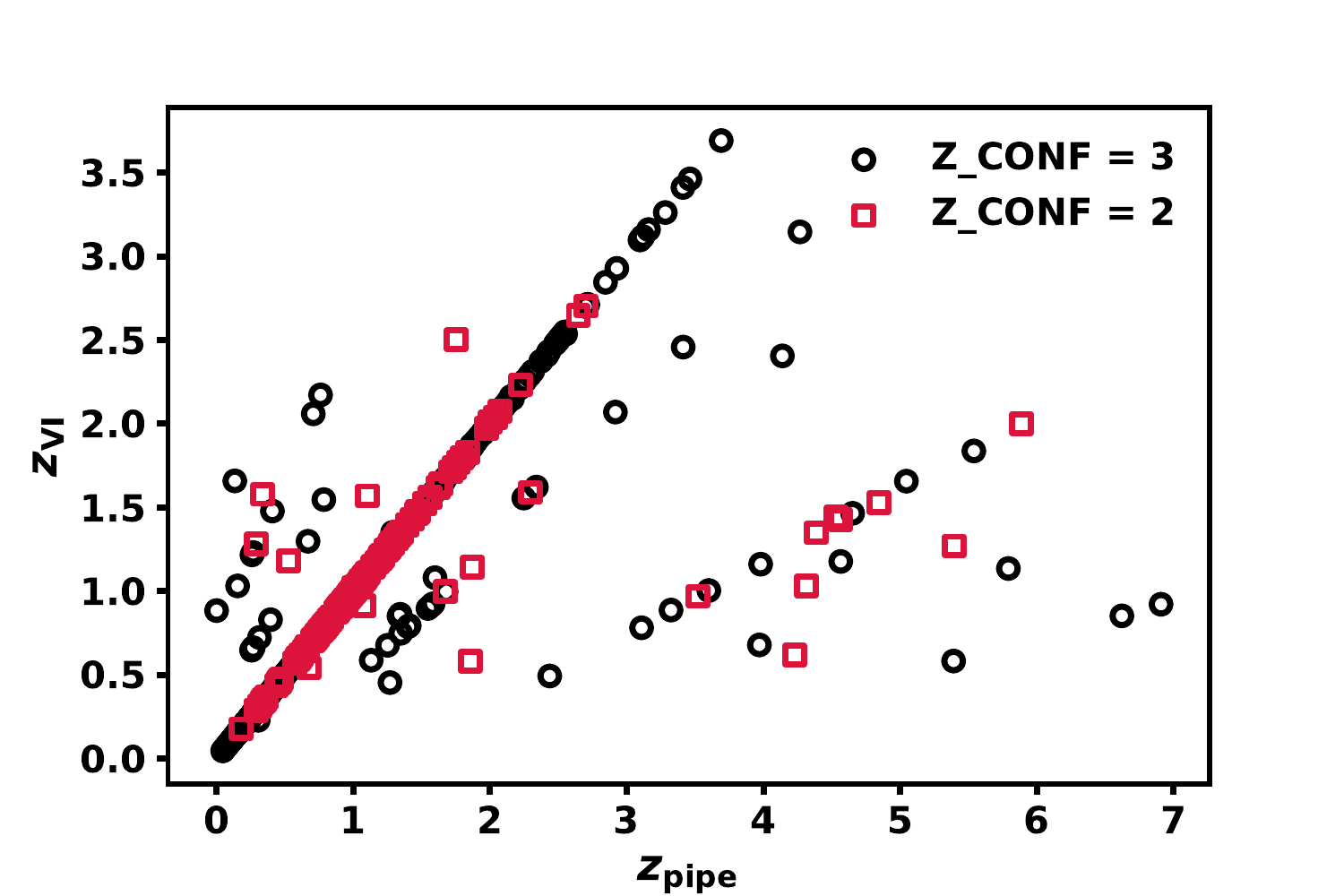}
  \caption{\label{zvi_v_zpipe} Comparison of pipeline redshifts ($z_{\rm pipe}$) and those verified or re-calculated via visual inspection of the SDSS spectra ($z_{\rm VI}$) for extragalactic sources whose redshifts could be determined. We find discrepant redshifts than those produced by the pipeline in 73 sources (69 extragalactic sources and 4 stars misclassified as QSOs or galaxies by the SDSS pipeline); an additional 54 sources had consistent redshifts between the pipeline and visual inspection, but different classifications. We highlight the sources with red boxes where the redshift identification is less confident (i.e., {\sc z\_conf} = 2): the outliers are not pre-dominantly the lower confidence redshifts.}
  \end{center}
\end{figure}

In addition to the 493 sources where we were unable to determine a reliable redshift, we find spectroscopic redshifts that are different from the pipeline value for 73 sources (see Figure \ref{zvi_v_zpipe}). Furthermore, for 54 sources where the pipeline redshift agrees with that from visual inspection, we found different spectroscopic classifications: either the pipeline failed to identify a weak broad emission line apparent by eye and labeled a source a ``Galaxy" instead of a ``QSO", or vice versa.

In total, we find 1642 sources (73\% of targets) whose pipeline produced spectroscopic redshifts and classifications were deemed reliable via visual inspection. When considering the subset of 167 sources with lower confidence on the visually inspected redshift (i.e., {\sc z\_conf} = 2), we find that only 18 objects have discrepant redshifts from the pipeline value.

A non-null value of the SDSS {\sc zwarning} flag indicates potential problems with the pipeline fit to the SDSS spectrum. In 595 cases, the {\sc zwarning} flag was set: 414 of the 493 sources where we were unable to determine a redshift had a non-null {\sc zwarning} value. About 30\% of the sources flagged by the {\sc zwarning} field did have spectra of sufficient quality to determine a redshift and classification. In 61 out of the 127 cases where we found a different redshift or spectroscopic classification than the pipeline, the {\sc zwarning} flag was also non-null.

\begin{figure}[ht]
  \begin{center}
  \includegraphics[scale=0.6]{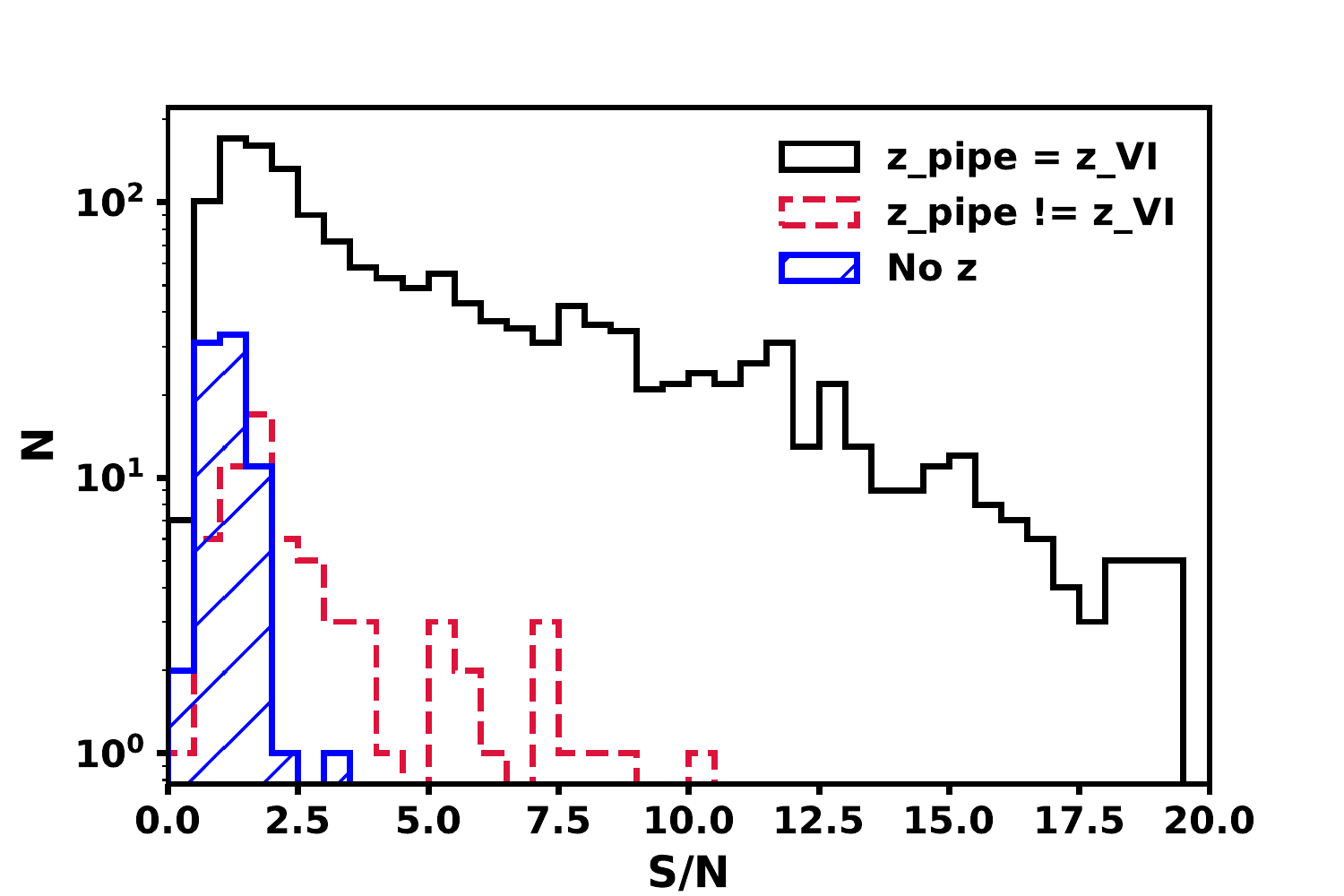}
  \caption{\label{sn_zwarn0} Signal-to-noise ratio (S/N) distribution of sources where the {\sc zwarning} flag was null. Most of the unidentifiable sources (blue hatched histogram) have S/N values below 2.25, as do $\sim$60\% of the sources where the pipeline redshift ($z_{\rm pipe}$) disagrees with that from visual inspection ($z_{\rm VI}$; red dashed histogram), suggesting that sources whose spectra have S/N below this threshold warrant visual inspection. The black solid line represents the distribution for sources where visual inspection deemed that the pipeline redshift and classification were reliable.}
  \end{center}
\end{figure}

Is there a way to immediately identify the remaining 145 sources where the pipeline redshift or classification was found to be unreliable via visual inspection, but the {\sc zwarning} flag did not indicate a potential error? We look at the S/N of the spectrum for clues. In Figure \ref{sn_zwarn0}, we plot the S/N for sources where the {\sc zwarning} flag was null for the following subsets: visual inspection confirmed the pipeline determined redshift and classification, we were able to determine a redshift from visual inspection that differed from that calculated by the pipeline, and  we were unable to measure a redshift from the spectrum. As expected, the sources where we were unable to determine a redshift have the lowest S/N spectrum, while the sources where visual inspection revealed a different redshift from the pipeline have a range of S/N values.

Our results indicate that in the absence of the automated {\sc zwarning} flag raising an alarm that the spectral fit may be problematic, the S/N can be used as a proxy. Seventy-eight of the 79 sources with spectra that were unidentifiable but had the {\sc zwarning} flag set to null have S/N values below 2.25. About 60\% of the sources (39 out of 66) where visual inspection revealed a different redshift than the pipeline are also below this S/N limit. This S/N cut could potentially be used to automatically reject any spectral classifications below this threshold. However 509 sources whose pipeline redshifts were deemed reliable via visual inspection (i.e., 29\% of identified sources) would be discarded with such an automatic cut.

To balance the competing demands of maximizing sample size with reliable spectral classifications and limited resources, we suggest that visual inspection of any source where {\sc zwarning} is non-null or S/N $<$ 2.25 would be prudent. Though most of the spectra will be unclassifiable when the {\sc zwarning} flag is set, about 30\% of the sources should be recoverable with visual inspection. About 20\% of sources that are not flagged as potentially problematic by the {\sc zwarning} output and have S/N below 2.25 are either unclassifiable or have different redshifts and/or classifications than the pipeline. Conversely, sources whose spectra are not flagged by {\sc zwarning} and have S/N $>$ 2.25 have largely reliable pipeline measurements: only 27 out of 1040 sources ($\sim$3\%) that meet these criteria have different redshifts/classification from the pipeline.

\subsection{Creating the Spectroscopic Sample}
As mentioned earlier, revised SDSS counterparts to the Stripe 82 X-ray sources \citep{ananna} and $W1-W2$ AGN color selection criteria \citep{assef2018} were published after the eBOSS Stripe 82X observations. To ensure where are using the most up-to-date information, with the most reliable counterparts and current MIR AGN definition, we only retain X-ray and {\it WISE} AGN targets that are marked as X-ray counterparts in the catalog of \citet{ananna} or {\it WISE} AGN candidates that obey the \citet{assef2018} $W1-W2$ color selection at the 75\% level. Our catalog is further vetted to only include eBOSS sources for which we were able to verify or independently determine a redshift ({\sc z\_conf} $\geq$ 2).


To create a complete spectroscopic catalog of X-ray sources and {\it WISE} AGN within this portion of the Stripe 82X survey, we include spectroscopic redshifts of SDSS counterparts to X-ray sources and {\it WISE} AGN from:
  \begin{itemize}
  \item Ancillary eBOSS Stripe 82X targets that are X-ray and {\it WISE} AGN counterparts based on these updated definitions (23 sources, for 1723 sources total from the SDSS-IV eBOSS Stripe 82X program);
    \item Sources whose spectra became available in SDSS DR14 but were not targeted as part of the SDSS-IV eBOSS Stripe 82X survey \citep[1670 sources;][]{sdssdr14,dr14q};
    \item Previous SDSS data releases whose {\it zwarning} flag is null \citep{albareti, paris, alam2015,sdssdr8,sdssdr7, ross2012} or whose spectra were independently vetted in a previous release of the Stripe 82X catalog \citep[1407 sources;][]{lamassa2017};
  \item 2SLAQ \cite[29 sources;][]{croom};
  \item 6dF \citep[2 sources;][]{jones2004,jones2009};
  \item Dedicated follow-up observing programs led by the members of the Stripe 82X collaboration (16 sources).
    \end{itemize}

  In total, our spectroscopic sample consists of 4847 sources, out of a parent sample of 10702 X-ray and {\it WISE} AGN candidates that lie within the SDSS-IV eBOSS Stripe 82X survey footprint. In the spectroscopic sample, we have 1891 X-ray sources, 3657 {\it WISE} sources, and 701 sources that are both.

Before discussing the completeness of the relative samples, we note that a subset of sources that lie along the North/South border of the SDSS scans within Stripe 82 lack photometry in the \citet{jiang} catalog (see their Figure 1). While creation of the eBOSS target list for the {\it WISE} AGN candidates was based on photometry from the \citet{jiang} catalog, we supplement this information with SDSS single-epoch photometry for both the spectroscopic sample and the parent sample. The sources with photometry from the SDSS single-epoch imaging is a small percentage of the total, amounting to 3.8\% of the spectroscopic sample and 5.2\% of the parent sample, respectively.

With the caveat in mind that we are using photometry from two different catalogs, we estimate the spectroscopic completeness of our samples based on the $r$-band magnitude. We highlight that 12.5\% of sources in the parent sample do not have photometric measurements in the $r$-band. Based on inspecting the magnitude distributions of these sources at other optical wavebands, we see that the $r$-band drop-outs are likely undetected as they are fainter than the $r$-band limit of the survey. We only consider the spectroscopic completeness for the subset of sources that are detected in the $r$-band, noting that this value is an upper limit for the full sample, but reasonable to the $r$-band depth of the \citet{jiang} catalog ($r \sim 24.6$).

\begin{figure*}[ht]
  \begin{center}
    \includegraphics[scale=0.4]{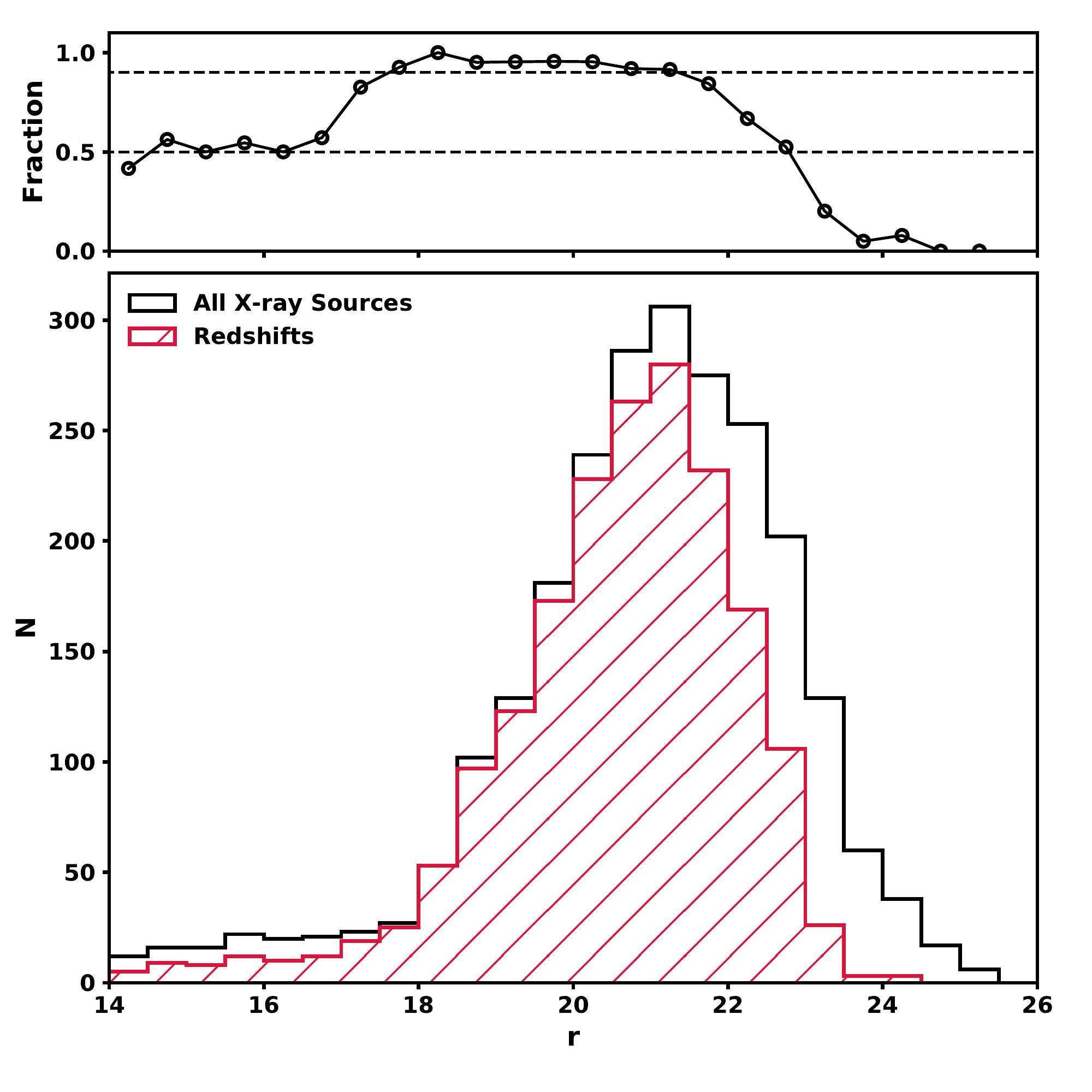}
    \includegraphics[scale=0.4]{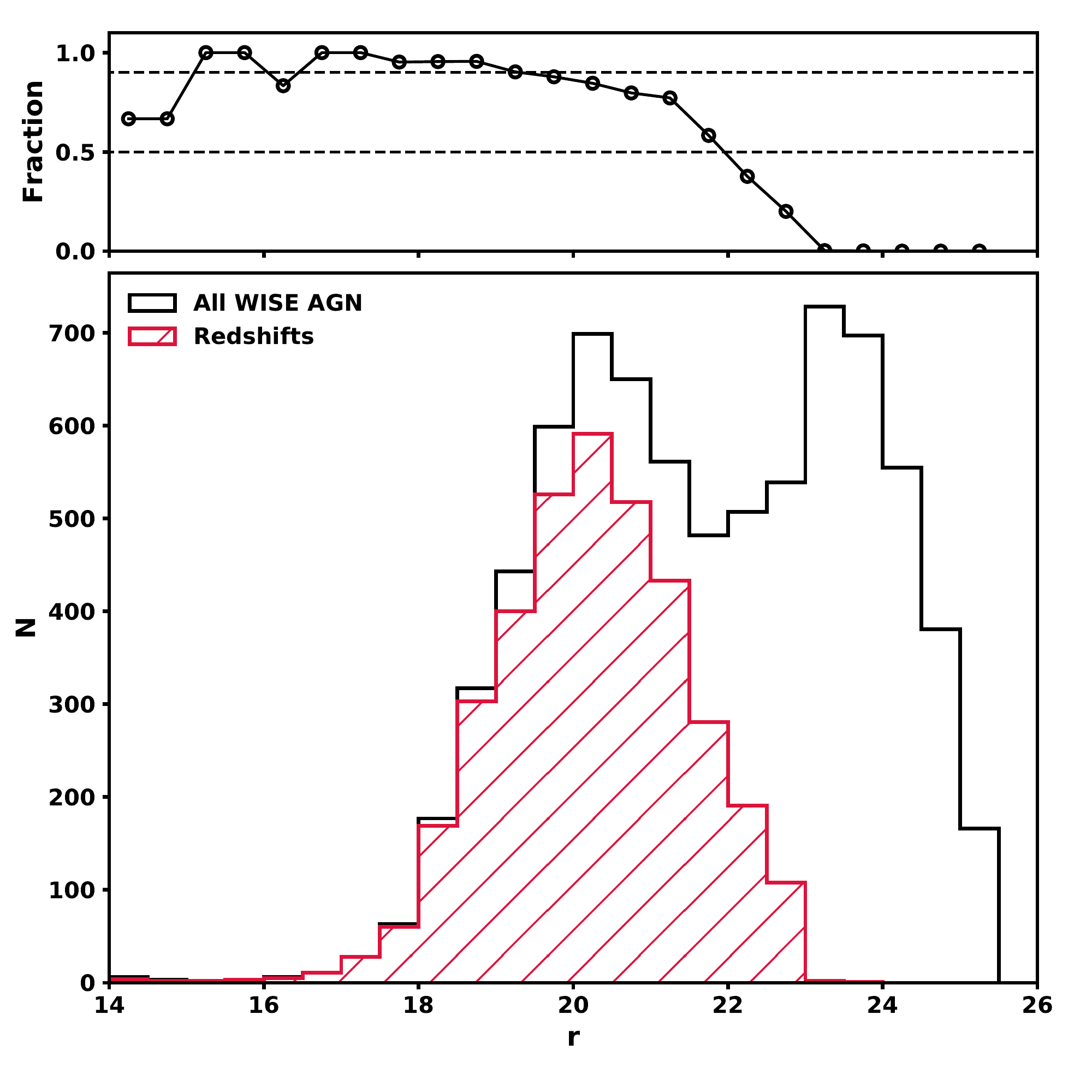}
  \caption{\label{spec_compl}$r$-band distribution of the parent sample of X-ray ({\it left}) and {\it WISE} sources ({\it right}) within the SDSS-IV eBOSS Stripe 82X survey area (black histogram), with those having spectroscopic redshifts shown by the red hatched histogram. The fraction of sources with spectroscopic identifications in the corresponding $r$-band magnitude bins is shown in the top plot, with horizontal lines indicating 90\% and 50\% completeness. Up to $r \sim 23$ (the magnitude limit of the eBOSS Stripe 82X survey) the X-ray sample is 82\% complete and the {\it WISE} AGN sample is 71\% complete. This completeness increases to 88\% and 82\% at $r \sim 22$ for the X-ray and {\it WISE} AGN sample, respectively.}
  \end{center}
\end{figure*}

In Figure \ref{spec_compl}, we show the number of X-ray sources and {\it WISE} AGN with spectroscopic redshifts compared with their parent samples as a function of $r$-band magnitude. We immediately see that the relatively low spectroscopic completeness of 45\% for the combined sample is due to the bi-modal distribution in the $r$-band magnitudes for the {\it WISE} sources, where there is an optically faint population that peaks at $r \sim 24$. When considering the samples separately, we find that the X-ray sample is 74\% complete while the {\it WISE} sample is 41\% complete. Considering the $r$-band limit of the eBOSS Stripe 82X survey ($r \sim 23$), the spectroscopic completeness rises to 72\% for the full sample, and 82\% and 71\% for the X-ray and {\it WISE} samples, respectively. At $r\sim 22$, the combined sample is 82\% complete, with the X-ray sample being 88\% complete and the {\it WISE} sample being 82\% complete.

\subsection{Description of Multi-Wavelength Information in Catalog}
In the catalog, we list the redshifts and optical spectroscopic classifications from the various surveys. The source of the spectroscopic redshift is noted in the column ``z\_src'' (see Table \ref{cat_cols}). If the redshift is from the SDSS-IV eBOSS Stripe 82X survey, the confidence on the vetted redshift is reported in ``z\_conf'', as described in Table \ref{vi_scheme}, otherwise it is set to a null value. The ``opt\_src'' column indicates whether the optical photometry is from the \citet{jiang} catalog or the SDSS pipeline. If the former, the magnitudes represent the ``AUTO'' magnitude from {\tt SExtractor} (i.e., Kron-like elliptical aperture). If the source is not detected in the \citet{jiang} catalog but has photometric measurements in the single-epoch SDSS catalog, then the optical magnitudes are the ModelMags from the SDSS pipeline, which for extended sources represent the better of an exponential profile fit or a  de Vaucouleurs profile fit, while a PSF model is used for point sources. The {\it WISE} magnitudes measured from profile-fitting photometry are are also reported, if the source is detected by {\it WISE}. We include the {\it WISE} photometry for every X-ray source detected by {\it WISE}, regardless of whether the source has $W1-W2$ AGN colors.

For the X-ray sources, we report the flux in the soft (0.5 - 2 keV), hard (2 - 10 keV for {\it XMM-Newton}; 2 - 8 keV for {\it Chandra}), and full (0.5 - 10 keV for {\it XMM-Newton}; 0.5 - 8 keV for {\it Chandra}) bands, as well as the significance of the detection in the corresponding ``soft\_detml,'' ``hard\_detml,'' and ``full\_detml'' columns \citep{lamassa2016a}, where $det\_ml$ = -ln$P_{\rm random}$, with $P_{\rm random}$ as the Poissonian probability that the detection is a random fluctuation. For the energy bands where $det\_ml \geq 10$ (i.e., $P_{\rm random} = 4.5\times10^{-5}$, 4$\sigma$ detection significance), we calculated the $k$-corrected X-ray luminosity\footnote{$L_{\rm k-corr} = L_{\rm observed} \times (1+z)^{\Gamma - 2}$, where $\Gamma$ is the powerlaw slope of the X-ray spectrum. For the Stripe 82X survey, we assumed $\Gamma$=1.7 for the hard and full bands and $\Gamma$=2.0 for the soft band \citep[see][]{lamassa2013b,lamassa2016a}.} for extragalactic sources. We emphasize that the reported X-ray sensitivity of the Stripe 82X survey is calculated for a higher significance value, namely for $det\_ml \geq 15$ (5.1$\sigma$) for the {\it XMM-Newton} observations and 4.5$\sigma$ for the archival {\it Chandra} observations.

From the X-ray fluxes, we calculated a hard-band X-ray luminosity ($L_{\rm X}$) which we use to classify whether a source is an X-ray AGN \citep[$L_{\rm X} > 10^{42}$ erg s$^{-1}$][]{brandt2005,brandt2015}. If the hard band X-ray flux is measured at $det\_ml \geq 10$, then we use this luminosity as $L_{\rm X}$. Otherwise, if the full band detection is significant at the $det\_ml \geq 10$ level, we scale the full band $k$-corrected luminosity by 0.665 to convert from the 0.5 - 10 keV band to the 2 - 10 keV band and estimate $L_{\rm X}$. If both the hard and full band detections are not significant at this level, then the soft band flux is scaled by a factor of 1.27 to convert from the 0.5 - 2 keV band to the 2 - 10 keV band for an estimate of $L_{\rm X}$.

In the catalog, we include the $W1 - W2$ colors for sources detected by {\it WISE}. We also note whether the source would be classified as an AGN at the 90\% (``WISE\_AGN\_90'') or 75\% (``WISE\_AGN\_75'') reliability level based on the criteria in \citet{assef2018}:
\begin{equation}
  W1-W2>\begin{cases}
    \alpha_R \ \text{exp}\{\beta_R (W2 - \gamma_R)^2\}, &  W2 > \gamma_R \\
    \alpha_R, & W2 \leq \gamma_R,
  \end{cases}
\end{equation}

\noindent where ($\alpha_{R}$,$\beta_R$,$\gamma_R$) = (0.650, 0.153, 13.86) for the 90\% reliability selection and ($\alpha_{R}$,$\beta_R$,$\gamma_R$) = (0.486, 0.092, 13.07) for the 75\% reliability selection. When commenting on demographics below, we consider any source that obeys the 75\% reliability selection as a {\it WISE} AGN.

Finally, we include a column that indicates the reddening of the source by calculating the optical to mid-infrared $R-W1$ color \citep{lamassa2016b}. We first convert the $r$-band magnitude in AB to $R$ in Bessel using:
\begin{equation}
  R({\rm AB}) = r - 0.0576 - 0.3718 \times ((r - i) - 0.2589),
\end{equation}
\citep{blanton}, then to Vega with:
\begin{equation}
  R({\rm Vega}) = R({\rm AB}) - 0.21
\end{equation}
As demonstrated in \citet{lamassa2016b}, obscured AGN between $0.5 < z < 1$ tend to have $R-W > 4$ colors, which is a redshift regime where the traditional BPT diagnostic becomes untenable for observed-frame optical spectra.

We perform BPT analysis on the low redshift sources ($z < 0.5$) that are spectroscopically classified as ``Galaxies'' but have X-ray luminosities or $W1 - W2$ colors consistent with AGN and relevant emission line fluxes with a S/N $>$ 5 (see below). Thus, we also report the ratios of [NII] 6584$\mathrm{\AA}$/H$\alpha$ and [OIII] 5007$\mathrm{\AA}$/H$\beta$ and the BPT classifications in the published catalog, where applicable.

All catalog columns are summarized in Table \ref{cat_cols}.

\clearpage
\begin{longtable*}{p{5cm}p{10cm}}
\tablecaption{\label{cat_cols} Stripe 82X eBOSS Value Added Catalog Column Descriptions}
\tablehead{\colhead{Column Name} & \colhead{Description}}

  SDSS RA & SDSS Right Ascension (J2000) \\

  SDSS Dec & SDSS Declination (J2000) \\

  MJD & Modified Julian Date of SDSS spectroscopic observation; only applicable to sources with SDSS spectroscopy \\

  Fiber & Fiber ID number of SDSS spectroscopic target; only applicable to sources with SDSS spectroscopy \\

  Plate & Plate number of SDSS spectroscopic observation; only applicable to sources with SDSS spectroscopy \\

  Redshift & Spectroscopic redshift. If spectrum was derived from the SDSS-IV eBOSS Stripe 82X program, it was vetted or independently determined by visual inspection. \\

  Class & Optical spectroscopic classification, vetted via visual inspection. Entries are: ``STAR'', ``QSO'' (if at least one broad emission line is present), ``GALAXY'' (only narrow emission lines or absorption lines are present). \\

   z\_src &  Source of spectroscopic redshift and classification:
    \begin{itemize}
    \item eBOSS\_S82X - eBOSS spectroscopic survey of Stripe 82X, described in this paper;
    \item SDSS\_DR14 - SDSS Data Release 14 \citep{sdssdr14};
    \item SDSS\_DR14Q - SDSS Quasar Catalog Data Release 14 \citep{dr14q};
    \item SDSS\_DR13 - SDSS Data Release 13 \citep{albareti};
    \item SDSS\_DR12 - SDSS Data Release 12 \citep{alam2015};
    \item SDSS\_DR12Q - SDSS Quasar Catalog Data Release 12 \citep{paris};
    \item SDSS\_DR8 - SDSS Data Release 8 \citep{sdssdr8};
    \item SDSS\_DR7Q - SDSS Quasar Catalog Data Release 7 \citep{sdssdr7};
    \item SDSS\_zwarning\_verified\_by\_eye - SDSS {\it zwarning} flag was non-null, but spectrum was vetted in the catalog release of \citet{lamassa2016a};
    \item pre-BOSS - pre-BOSS pilot survey using Hectospec on MMT \citep{ross2012};
    \item 2SLAQ - 2SLAQ survey \citep{croom};
    \item 6dF - 6dF survey \citep{jones2004, jones2009};
    \item HYDRA\_2014\_Jan - follow-up observations of Stripe 82X sources from WIYN HYDRA in 2014 Jan, first published in \citet{lamassa2016a};
    \item HYDRA\_2015\_Jan - follow-up observtions of Stripe 82X sources from WIYN HYDRA  in 2015 Jan, first published in \citet{lamassa2016a};
    \item Gemini\_GNIRS\_2015 - follow up observations of obscured AGN candidates in Stripe 82X from Gemini GNIRS in 2015, first published in \citet{lamassa2017};
    \item DBSP\_2015\_Sep - follow-up observations of Stripe 82X sources from Palomar DoubleSpec in 2015 Sep, published here for the first time;
    \item DBSP\_2016\_Aug - follow-up observations of Stripe 82X sources from Palomar DoubleSpec in 2016 Aug, published here for the first time;
    \item DBSP\_2017\_Oct - follow-up observations of Stripe 82X sources from Palomar DoubleSpec in 2017 Oct, published here for the first time;
    \item Keck\_LRIS\_Oct2017 - follow-up observations of Stripe 82X sources from Keck LRIS in 2017 Oct, published here for the first time.
  \end{itemize} \\
      
  z\_conf & Confidence on spectroscopic redshift via visual inspection. 2: one emission/absorption line identified, 3: $\geq$2 emission/absorption lines identified; only applicable to sources from the SDSS-IV eBOSS Stripe 82X program. \\

  u\_mag & {\tt SExtractor} ``AUTO'' (i.e., Kron-line elliptical aperture) $u$-band magnitude from coadded \citet{jiang} catalog (AB) or SDSS ModelMag photometric measurement. \\

  u\_err & error on $u$-band magnitude from coadded \citet{jiang} catalog or SDSS ModelMagErr value from SDSS pipeline. \\

  g\_mag & {\tt SExtractor} ``AUTO'' (i.e., Kron-line elliptical aperture) $g$-band magnitude from coadded \citet{jiang} catalog (AB) or SDSS ModelMag photometric measurement.\\

  g\_err & error on $g$-band magnitude from coadded \citet{jiang} catalog or SDSS ModelMagErr value from SDSS pipeline. \\

  r\_mag & {\tt SExtractor} ``AUTO'' (i.e., Kron-line elliptical aperture) $r$-band magnitude from coadded \citet{jiang} catalog (AB) or SDSS ModelMag photometric measurement. \\

  r\_err & error on $r$-band magnitude from coadded \citet{jiang} catalog or SDSS ModelMagErr value from SDSS pipeline. \\

  i\_mag &{\tt SExtractor} ``AUTO'' (i.e., Kron-line elliptical aperture)  $i$-band magnitude from coadded \citet{jiang} catalog (AB) or SDSS ModelMag photometric measurement. \\

  i\_err & error on $i$-band magnitude from coadded \citet{jiang} catalog or SDSS ModelMagErr value from SDSS pipeline. \\

  z\_mag & {\tt SExtractor} ``AUTO'' (i.e., Kron-line elliptical aperture) $z$-band magnitude from coadded \citet{jiang} catalog (AB) or SDSS ModelMag photometric measurement. \\

  z\_err & error on $z$-band magnitude from coadded \citet{jiang} catalog or SDSS ModelMagErr value from SDSS pipeline. \\

    opt\_src &  Source of optical photometry. J14 - coadded catalog of \citet{jiang}, SDSS - pipeline photometry from the single-epoch SDSS catalog. \\

  W1 & {\it WISE} magnitude at 3.4$\mu$m measured with profile-fitting photometry (Vega).  Only reported if W1 magnitude has a S/N $\geq$ 2. \\

  W1sig & uncertainty on $W1$ magnitude \\

  W2 & {\it WISE} magnitude at 4.6$\mu$m measured with profile-fitting photometry (Vega). Only reported if W2 magnitude has a S/N $\geq$ 2.  \\

  W2sig & uncertainty on $W2$ magnitude \\

  W3 & {\it WISE} magnitude at 12$\mu$m measured with profile-fitting photometry (Vega). Only reported if W3 magnitude has a S/N $\geq$ 2.  \\

  W3sig & uncertainty on $W3$ magnitude \\

  W4 & {\it WISE} magnitude at 22$\mu$m measured with profile-fitting photometry (Vega).  Only reported if W4 magnitude has a S/N $\geq$ 2.  \\

  W4sig & uncertainty on $W4$ magnitude. \\

  R-W1 & $R-W1$ (Vega) color, useful to assess reddening of source. \\

  $W1-W2$  & {\it WISE} $W1-W2$ color, which is used to determine whether the source is a {\it WISE} AGN. \\

  WISE\_AGN\_75 & If {\it WISE} $W1-W2$ color obeys the \citet{assef2018} AGN selection at the 75\% reliability threshold, this flag is set to ``YES.'' In the main body of the text, a source that meets this color criterion is considered a {\it WISE} AGN. \\

    WISE\_AGN\_90 & If {\it WISE} $W1-W2$ color obeys the \citet{assef2018} AGN selection at the 90\% reliabilibty threshold, this flag is set to ``YES''\\

  Soft\_flux & X-ray flux in the 0.5-2 keV band from \citet{lamassa2016a} \\

  Soft\_DETML & Significance of X-ray detection in the soft band, where $det\_ml$ = -ln$P_{\rm random}$, with $P_{\rm random}$ as the Poissonian probability that the detection is a random fluctuation. \\

  Soft\_Lum & Log of the $k$-corrected soft-band X-ray luminosity. Only computed if Soft\_DETML $>$ 10.\\

  Hard\_flux & X-ray flux in the 2-10 (2-7) keV band for {\it XMM-Newton} ({\it Chandra}) sources from \citet{lamassa2016a} \\

  Hard\_DETML & Significance of X-ray detection in the hard band, where $det\_ml$ = -ln$P_{\rm random}$, with $P_{\rm random}$ as the Poissonian probability that the detection is a random fluctuation. \\

  Hard\_Lum & Log of the $k$-corrected hard-band X-ray luminosity. Only computed if Hard\_DETML $>$ 10.\\

  Full\_flux & X-ray flux in the 0.5-10 (0.5-7) keV band for {\it XMM-Newton} ({\it Chandra}) sources from \citet{lamassa2016a} \\

  Full\_DETML & Significance of X-ray detection in the full band, where $det\_ml$ = -ln$P_{\rm random}$, with $P_{\rm random}$ as the Poissonian probability that the detection is a random fluctuation. \\

  Full\_Lum & Log of the $k$-corrected full-band X-ray luminosity. Only computed if Full\_DETML $>$ 10.\\

  Xray\_Lum & Estimate of the $k$-corrected 2-10 keV X-ray luminosity ($L_{\rm X}$). If Hard\_DETML $>$ 10, this value respresents the measured $k$-corrected hard X-ray luminosity. Otherwise, if Full\_DETML $>$ 10, the $k$-corrected full-band luminosity is adjusted by a factor of 0.665 to convert to the 2-10 keV band luminosity. If both Hard\_DETML and Full\_DETML are below 10, the $k$-corrected soft-band luminosity is scaled by a factor of 1.27 to convert to the 2-10 keV band luminosity. \\

  Xray\_AGN & If $L_{\rm X} > 10^{42}$ erg s$^{-1}$, the source is considered an X-ray AGN and this flag is set to ``YES'' \\
  
  Log(NII\_6584/H\_alpha) & Logarithm of the [NII] 6584$\mathrm{\AA}$/H$\alpha$ ratio. Only populated for $z < 0.5$ sources spectroscopically identified as ``Galaxies'' in the SDSS pipeline with a S/N $>$5 in the [NII] 6584 \AA, [OIII] 5007 \AA, H$\alpha$, and H$\beta$ lines. Emission line fluxes are measured by the SDSS pipeline. \\
  
  Log(OIII\_5700/H\_beta) & Logarithm of the [OIII] 5007$\mathrm{\AA}$/H$\alpha$ ratio. Only populated for $z < 0.5$ sources spectroscopically identified as ``Galaxies'' in the SDSS pipeline with a S/N $>$5 in the [NII] 6584 \AA, [OIII] 5007 \AA, H$\alpha$, and H$\beta$ lines. Emission line fluxes are measured by the SDSS pipeline. \\

  BPT Classification & BPT classification of sources at $z<0.5$ that are classified as ``Galaxies'' in the SDSS pipeline, and with emission line fluxes significant at the 5$\sigma$ level. ``Sy2'': Seyfert 2 galaxy based on the definition of \citet{kewley}; ``Comp'': Composite galaxy with emission line rations between the theoretical starburst line of \citet{kewley} and empirical dividing line of \citet{kauffmann}; ``SF'': Star-forming galaxies with emission line ratios below the \citet{kauffmann} demarcation.
  
\end{longtable*}

\section{Discussion}
 We report on the demographics of the AGN in the spectroscopic sample, dividing AGN based on their optical spectroscopic classifications:
\begin{itemize}
\item ``Type 1 AGN'' have at least one broad emission line in their SDSS spectra (i.e., labeled as ``QSO"s in the SDSS pipeline and Table \ref{vi_scheme});
\item ``optically obscured AGN'' have no broad lines (i.e., labeled as ``Galaxy" in Table \ref{vi_scheme}), but are AGN on the basis of their X-ray luminosity or {\it WISE} colors.
\end{itemize}
We note that four objects from archival spectroscopic databases do not report optical classifications, so we do not include these sources when discussing Type 1 versus optically obscured AGN. All four sources have extragalactic redshifts, one is an X-ray AGN, and one is a {\it WISE} AGN. In the Appendix, we discuss how the demographics of the Stripe 82 X-ray AGN compare with the AGN from the XMM-XXL Northern survey \citep{pierre2016,liu2016,menzel}.

\subsection{Demographics of X-ray and {\it WISE} AGN}
Of the 4847 sources in our spectroscopic sample, 4782 are extragalactic and 65 are stars (46 X-ray emitting stars and 19 stars with {\it WISE} $W1-W2$ colors that meet the AGN 75\% reliability threshold). Considering the extragalactic sample, we find 1790 X-ray AGN (i.e.,  $L_{\rm x} > 10^{42}$ erg s$^{-1}$), 3638 {\it WISE} AGN, and 698 X-ray and {\it WISE} AGN. Fifty-five sources are X-ray galaxies (i.e., $L_{\rm x} < 10^{42}$ erg s$^{-1}$), but three of these are classified as AGN based on their {\it WISE} $W1-W2$ colors.

In Table \ref{dem_summary}, we summarize the X-ray and {\it WISE} AGN classifications for the spectroscopic sample within the SDSS-IV eBOSS Stripe 82X field. Of the 4730 AGN in the survey area, 70\% are Type 1 and 30\% are optically obscured. As our sample is over 80\% complete to $r \sim 22$, these demographics may be representative of the X-ray and MIR-selected AGN population up to the optical, X-ray, and MIR flux limits of these surveys, but as we discuss further below, the total number of AGN, and number of optically obscured AGN, may be biased by a significant fraction of star-forming galaxy interlopers with {\it WISE} AGN colors at $z\sim 0.3$. When considering the {\it WISE} and AGN samples separately, we do fine a higher percentage of optically obscured {\it WISE} AGN (30\%) than X-ray AGN (20\%).

For reference, we include a demographic summary of the AGN classified from the SDSS-IV eBOSS Stripe 82X survey in Table \ref{dem_summary}, highlighting that an overwhelming fraction of the optically obscured AGN in the field (76\%) were garnered from this dedicated follow-up program. This results underscores the importance of spending resources to follow-up X-ray and MIR-selected AGN candidates for a complete census of obscured black hole growth.

\begin{deluxetable*}{lrrrr}
  \tablewidth{0pt}
  \tablecaption{\label{dem_summary} Demographic Summary of Stripe 82X Sources in SDSS-IV eBOSS Survey Area\tablenotemark{1}}
  \tablehead{\colhead{Classification} & \colhead{X-ray AGN\tablenotemark{2}} & \colhead{{\it WISE} AGN} & \colhead{X-ray \& {\it WISE} AGN } &
  \colhead{Total X-ray or {\it WISE} AGN}}

  \multicolumn{5}{c}{Total}\\
  \hline \\
  Type 1 AGN              & 1427    & 2529 & 646 & 3310 \\
  Optically Obscured AGN  & 362     & 1108 &  52 & 1418 \\
  Stars                   & 46      & 19   &   0 &   65\\

  \hline \\
  \multicolumn{5}{c}{SDSS-IV eBOSS Sources}\\
  \hline \\
  Type 1 AGN             &  335      & 354 & 85 &  604 \\
  Optically Obscured AGN &  269      & 844 & 40 & 1073 \\
  Stars                  &  27       & 13  &  0 &   40 

  \enddata
  \tablenotetext{1}{X-ray sources are mostly within the 15.6 deg$^2$ footprint of the {\it XMM-Newton} AO13 survey, while {\it WISE} sources are from the larger 36.8 deg$^2$ footprint of the SDSS-IV eBOSS Stripe 82X spectroscopic survey. We note that 1 X-ray AGN and 1 {\it WISE} AGN do not have spectroscopic classifications in archival databases and are thus not included in the Type 1 and optically obscured AGN census above.}
   \tablenotetext{2}{X-ray sources where $L_{\rm X} > 10^{42}$ erg s$^{-1}$ and are thus classifiable as AGN based on their powerful X-ray emission. We list the number of X-ray detected stars (and stars with {\it WISE} $W1-W2$ colors consistent with AGN) for reference. In addition to the sources listed here, there are 55 X-ray galaxies (i.e., X-ray sources with $L_{\rm X} < 10^{42}$ erg s$^{-1}$), 3 of which are classified as AGN based on their {\it WISE} colors; 6 of these X-ray galaxies are from the eBOSS survey.}
\end{deluxetable*}

\begin{figure*}[ht]
  \begin{center}
  \includegraphics[scale=0.55]{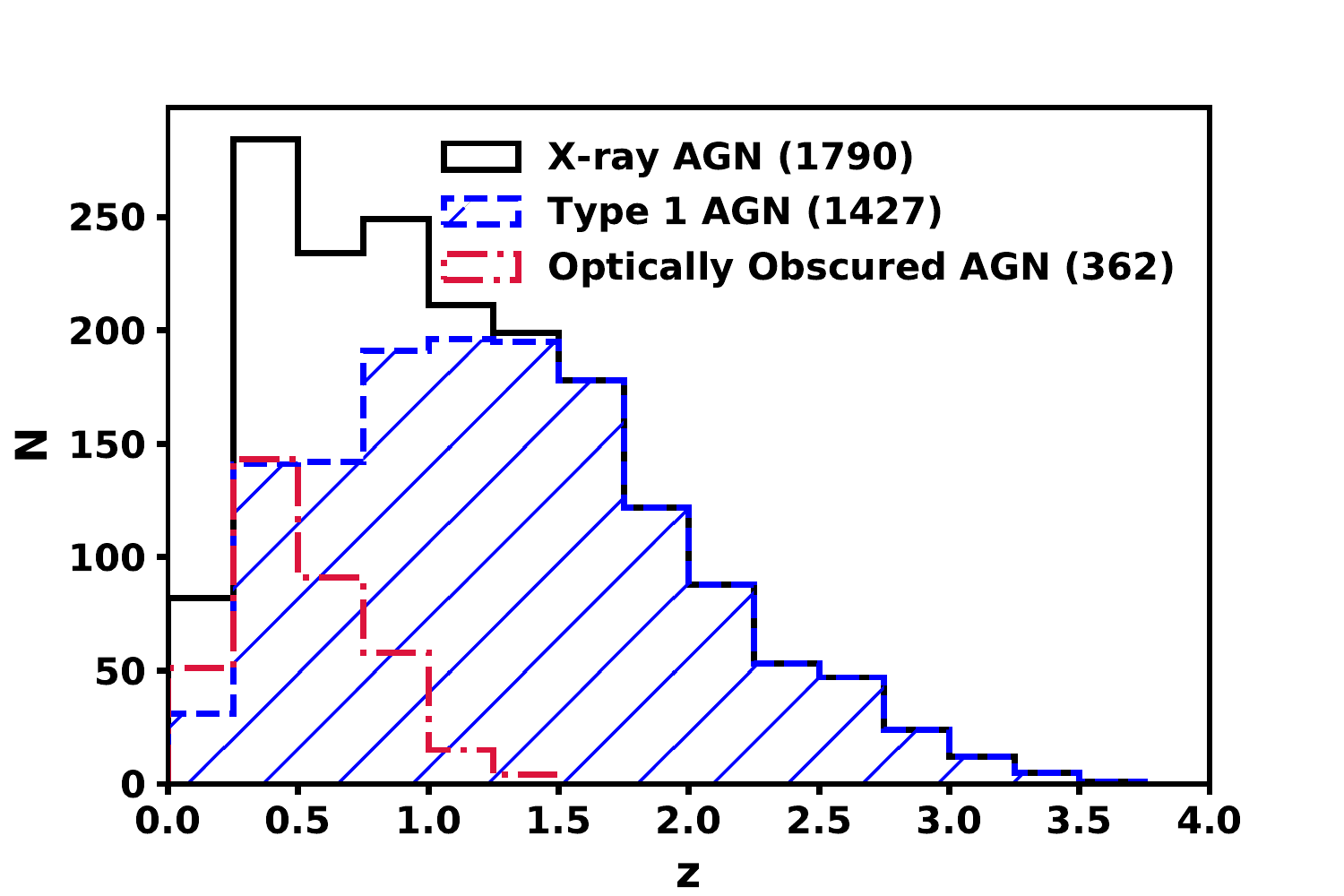}
  \includegraphics[scale=0.55]{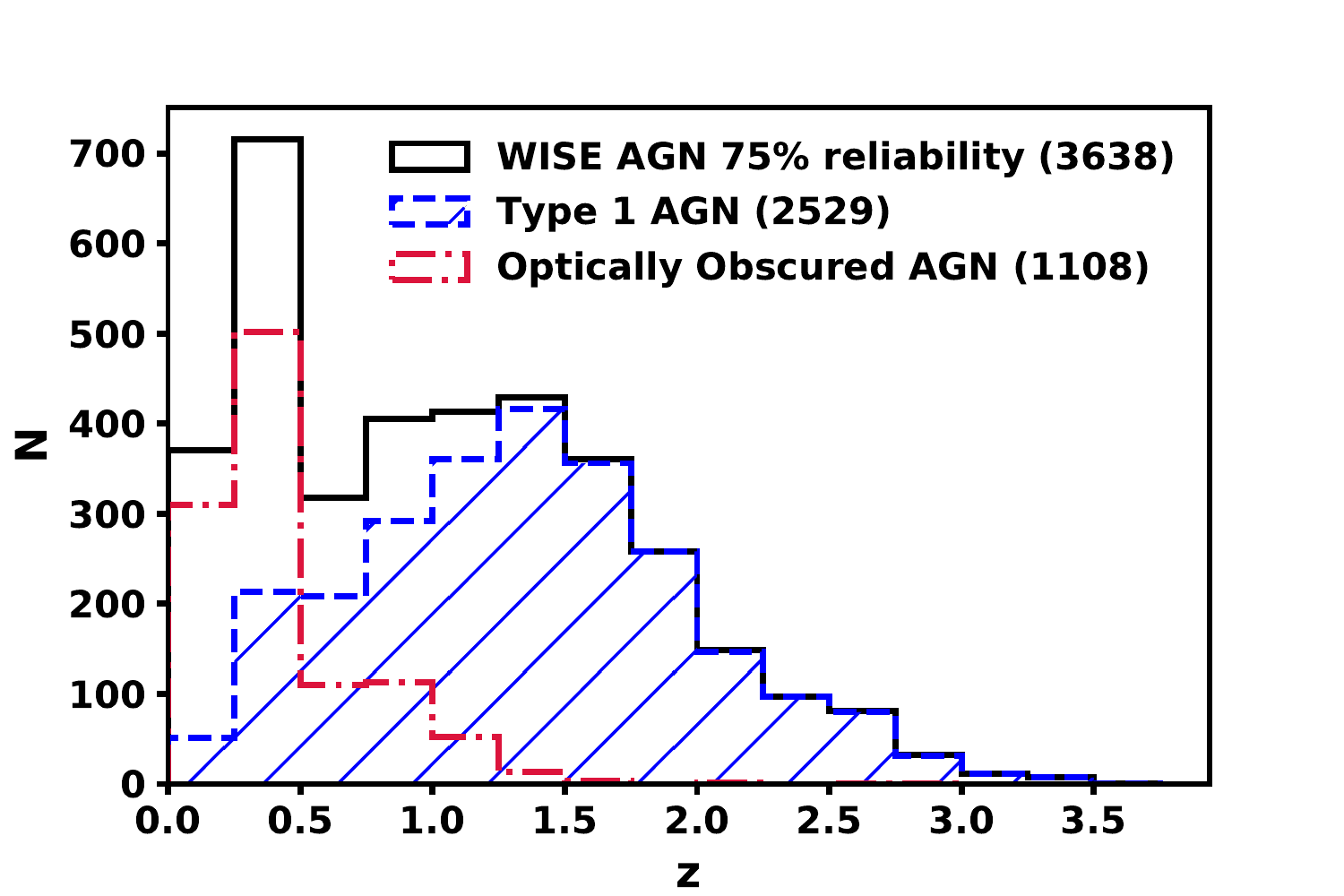}
  \caption{\label{z_distr} Redshift distribution of the ({\it left}) X-ray AGN and ({\it right}) {\it WISE} AGN; note that one X-ray and {\it WISE} AGN included in the total sample lack spectroscopic classifications in archival catalogs and are thus not included in the Type 1 and optically obscured AGN sub-samples. The subset of sources that are Type 1 AGN and optically obscured AGN are shown by the dashed blue, hatched histogram and the red dot-dashed histogram, respectively. A prominent peak at $z \sim 0.3$ is apparent in the {\it WISE} AGN optically obscured population. For both samples, the optically obscured AGN are at lower redshift than the Type 1 AGN. The median redshift of this sample of AGN is $z \sim 1$ for both populations.}
  \end{center}
\end{figure*}

The AGN span a redshift range of $0.02 < z < 4.2$, with a median redshift of $z\sim 1$, as shown in Figure \ref{z_distr}. In the redshift distribution for the optically obscured {\it WISE} AGN, a prominent peak is visible at $z \sim 0.3$. For both the X-ray and {\it WISE} AGN, the Type 1 sub-populations are visible to a higher redshift than the optically obscured AGN.

\subsubsection{BPT Analysis of Local Obscured AGN}
We perform BPT analysis for the subset of sources optically spectroscopically identified as ``Galaxies'' at $z < 0.5$ that have a signal-to-noise ratio of at least 5 in the H$\alpha$, H$\beta$, [OIII] 5007 \AA, and [NII] 6584 \AA\ lines. Here, we use the \citet{kewley} maximal starburst line to define Seyfert 2 galaxies, and the empirical \citet{kauffmann} demarcation to separate star-forming galaxies from composite galaxies, which have a mixture of star-forming and AGN ionization powering their emission.

\begin{figure*}[ht]
  \begin{center}
  \includegraphics[scale=0.55]{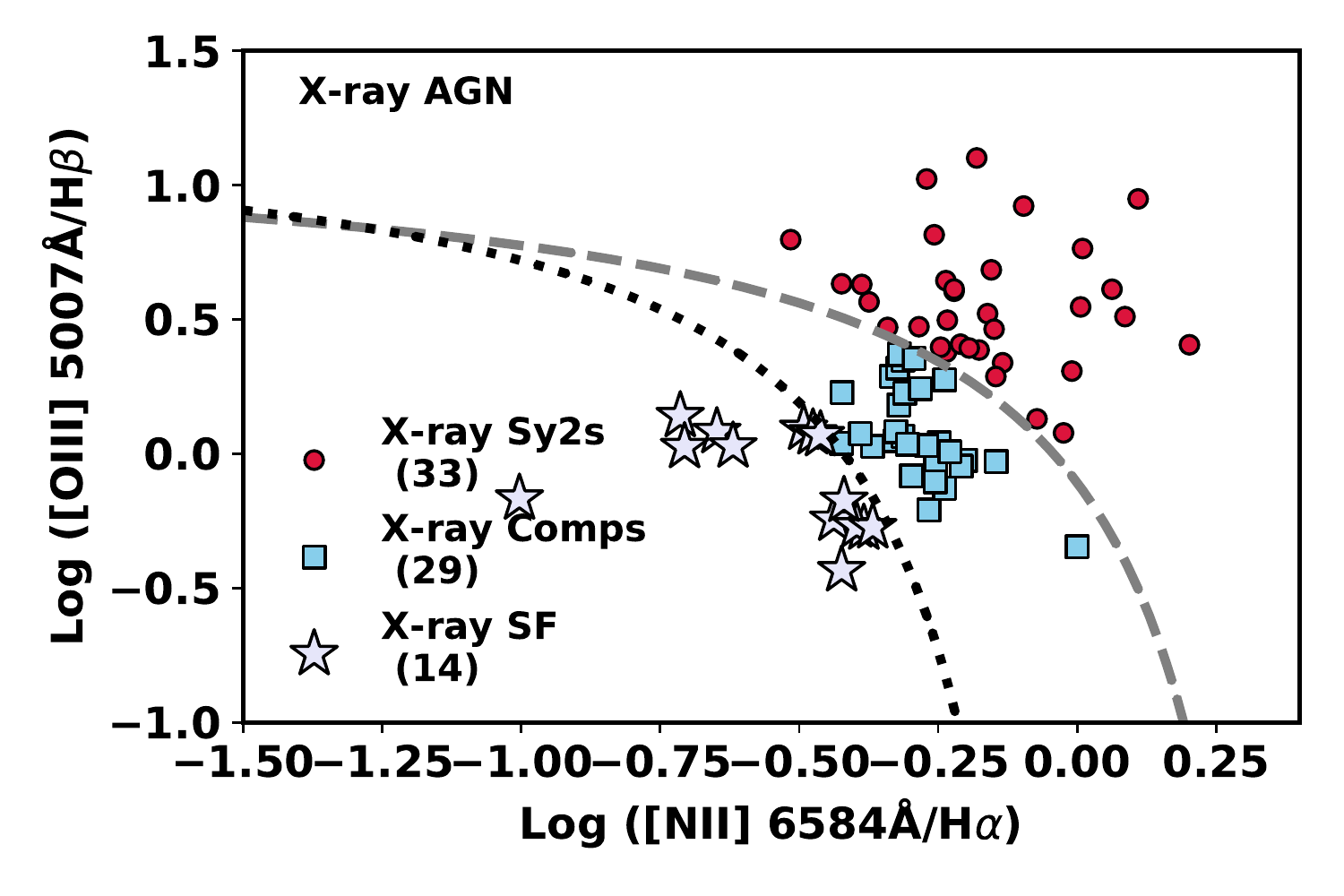}
  \includegraphics[scale=0.55]{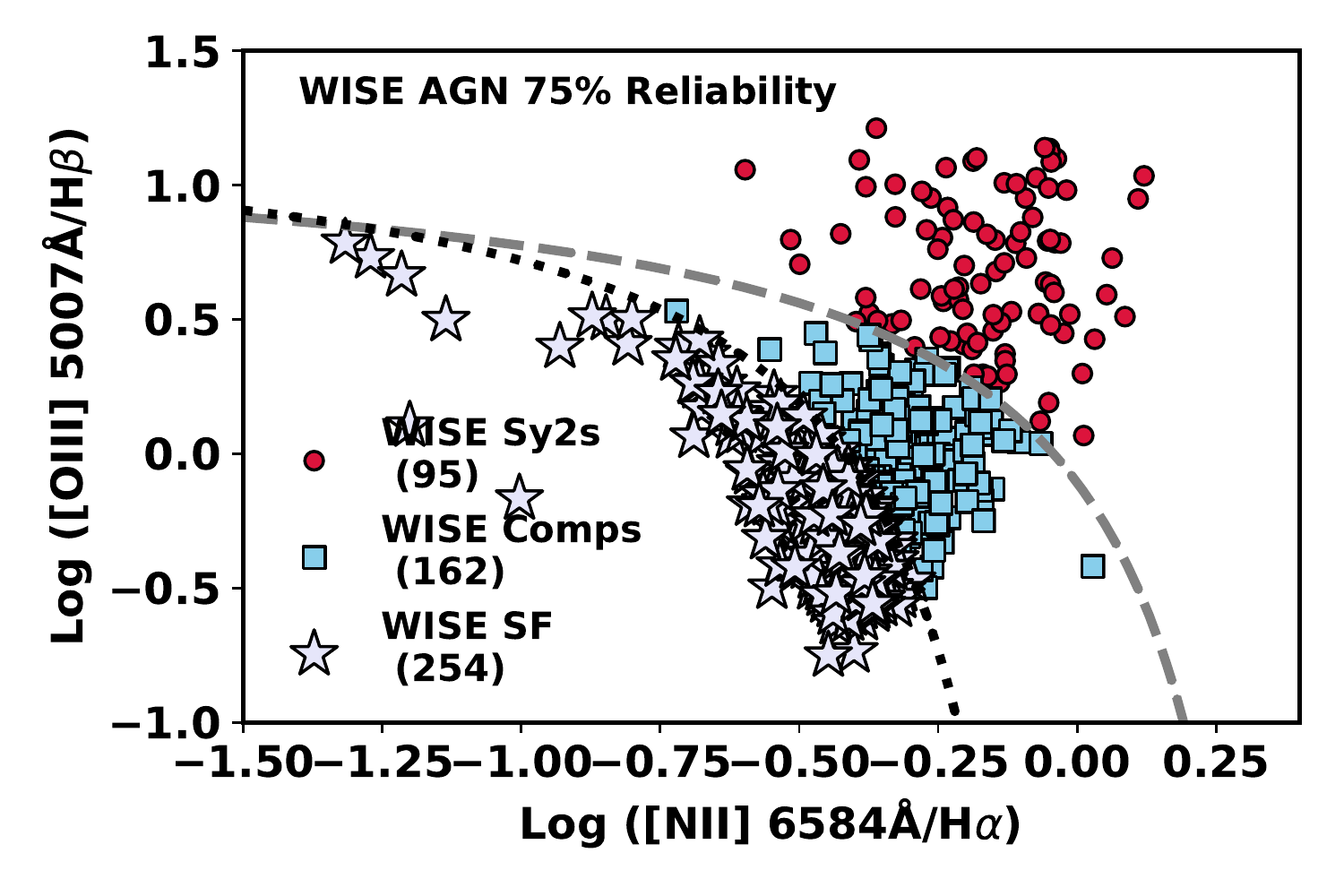}
  \caption{\label{bpt_all} BPT diagram for the $z < 0.5$ ({\it left}) X-ray AGN and ({\it right}) {\it WISE} AGN at the 75\% reliability level that are spectroscopically identified as ``Galaxies" and have a S/N exceeding 5 in the H$\alpha$, H$\beta$, [OIII] 5007 \AA, and [NII] 6584 \AA\ emission lines; fluxes are derived from the SDSS pipeline fit to the spectra. The grey dashed line notes the maximal starburst line from \citet{kewley} where the emission line ratios indicate AGN ionization and the black dotted line notes the empirical dividing line between star-forming and composite galaxies from \citet{kauffmann}. About a fifth of the X-ray AGN would be missed based on optical characteristics, while half of {\it WISE} AGN are spectroscopically identified as star-forming galaxies. Whether this latter class represents elusive AGN whose optical signatures are buried by dust or star-forming galaxies whose MIR colors mimic those of AGN is unclear. }
  \end{center}
\end{figure*}

For the X-ray AGN (left panel of Figure \ref{bpt_all}), we see that 18\% of the sources (14 out of 76) would be misclassified as non-AGN on the basis of their optical emission alone. However, based on their X-ray luminosities, these galaxies do host active central black holes. Similar results, i.e., X-ray AGN hosted in BPT-classified star-forming galaxies have been observed in other X-ray samples \citep[so-called ``optically elusive AGN'';][]{maiolino,caccianiga,menzel,pons,smith}. 

\begin{figure*}[ht]
  \begin{center}
  \includegraphics[scale=0.55]{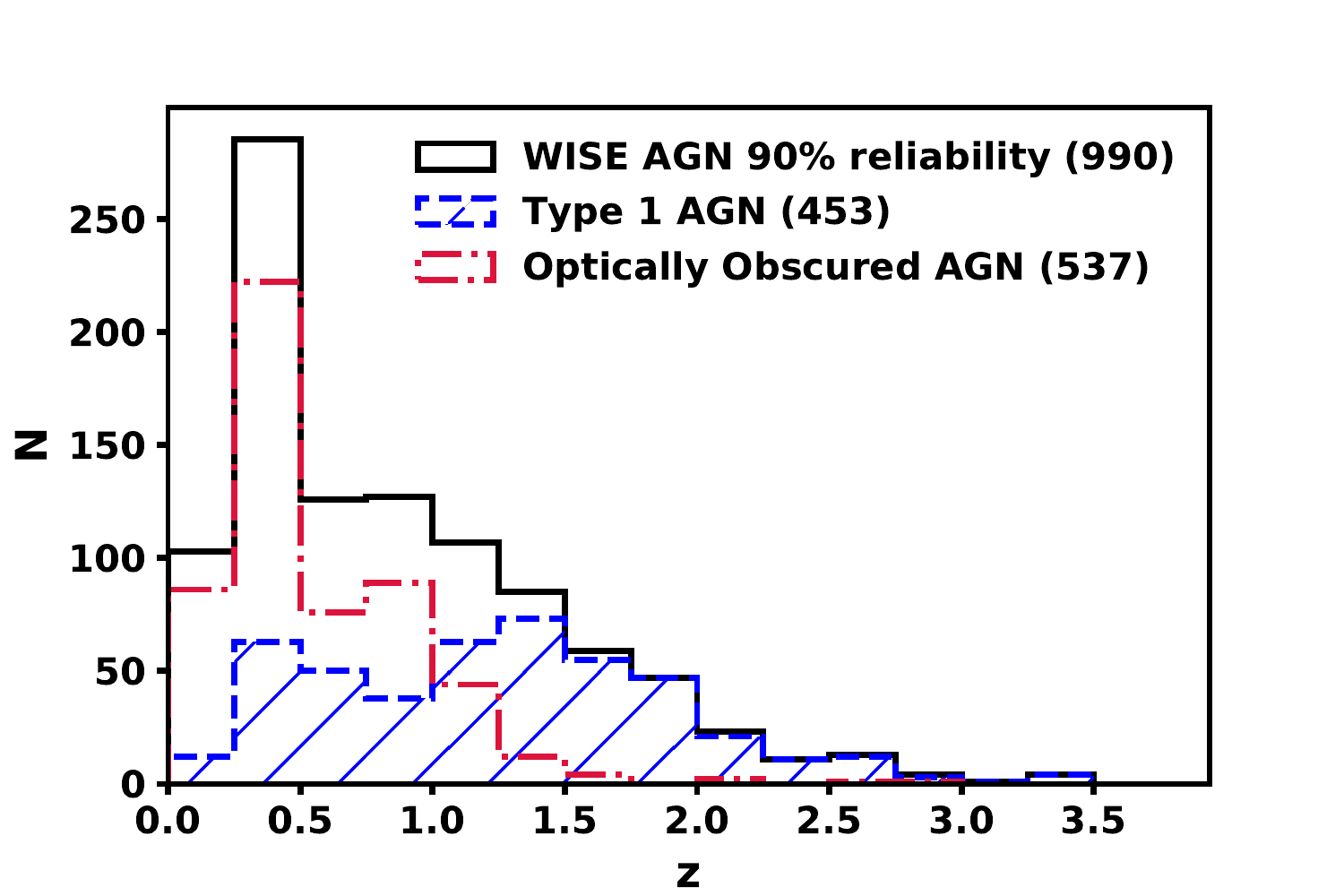}
  \includegraphics[scale=0.55]{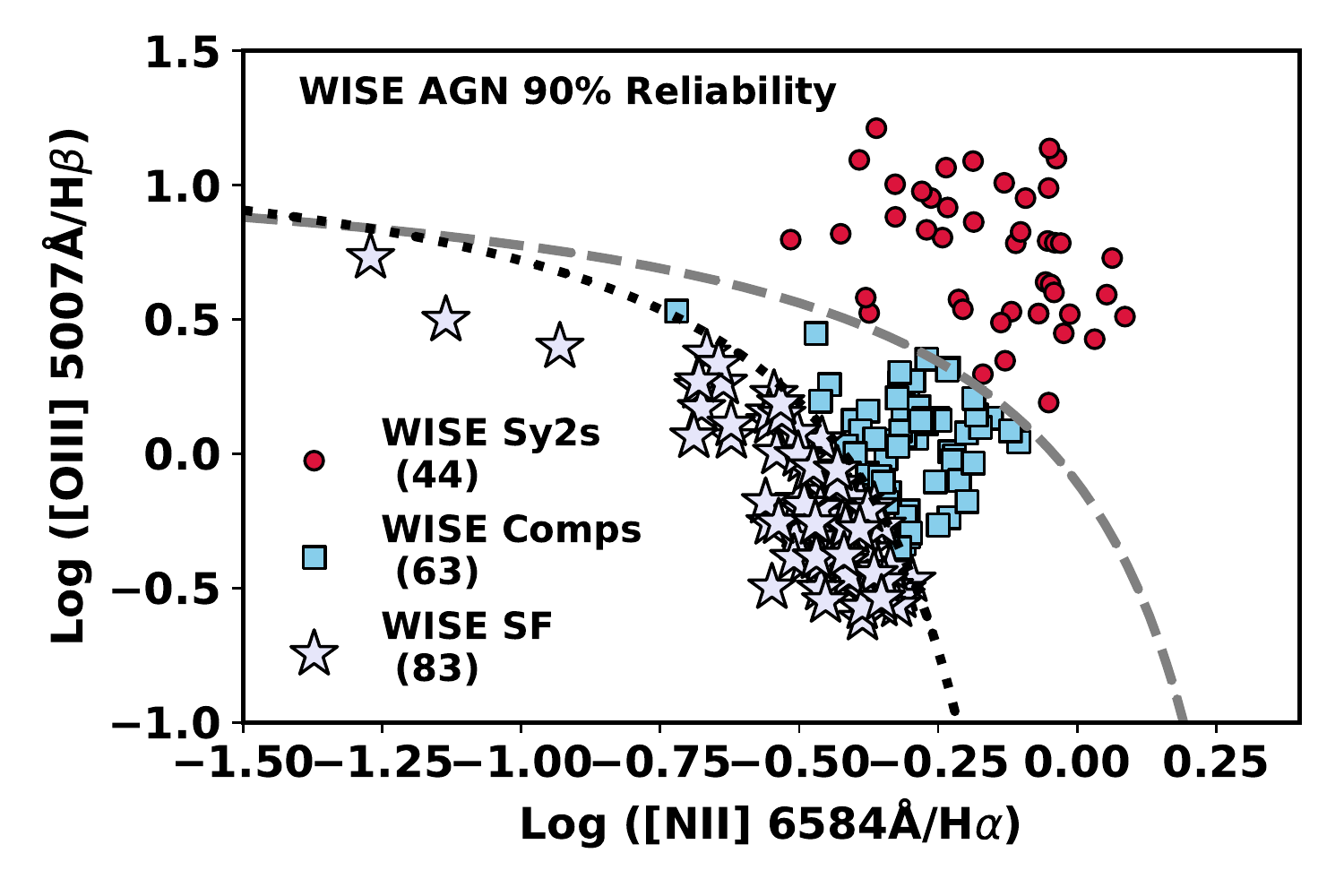}
  \caption{\label{wagn_90} {\it Left}: Redshift distribution and ({\it right}) BPT diagram of {\it WISE} AGN identified at the 90\% reliability level \citep{assef2018}. The prominent peak at $z \sim 0.3$ and relatively high percentage of {\it WISE} AGN spectroscopically classified as star-forming galaxies remain, indicating that more stringent MIR color cuts do not erase the trends observed at the 75\% reliability level. Theoretical modeling would be needed to test whether the {\it WISE} AGN color selection fails at $z\sim 0.3$, preferentially selecting non-active galaxies, or if this population represents a new class of buried AGN. }
  \end{center}
\end{figure*}

A much higher percentage (50\%) of {\it WISE} AGN at $z < 0.5$ are classified as star-forming galaxies (right panel of Figure \ref{bpt_all}). The nature of these objects is less clear than for the X-ray AGN. Combined with the redshift peak of {\it WISE} AGN at $z \sim 0.3$, these results may point to a degeneracy in {\it WISE} AGN color selection, where star-forming galaxies at these redshifts can have mid-infrared colors mimicking AGN \citep[see][]{satyapal}. If we restrict the {\it WISE} AGN to those defined at the 90\% reliability level of \citep{assef2018}, we see a similar trend: a peak in the AGN distribution at $z \sim 0.3$ remains (Figure \ref{wagn_90}, left), and the fraction of {\it WISE} AGN in the star-forming locus of the BPT diagram is similar (44\%, Figure \ref{wagn_90}, right). Distinguishing between optically buried AGN and star-forming galaxy impostors masquerading as MIR AGN at $z \sim 0.3$ would require further theoretical modeling \citep{satyapal}, which will be the topic of a future paper.

\subsection{Comparison between X-ray and {\it WISE} AGN}
Here we explore the characteristics of AGN found, and missed, by X-ray and MIR selection. We reiterate that the area of the SDSS plates in the eBOSS program is larger than the field of view of the {\it XMM-Newton} AO13 observations in Stripe 82 (Figure \ref{fig:survey_layout}). Thus, for the most straightforward comparison between the demographics of the X-ray and {\it WISE} AGN, we cull the {\it WISE} list to only include those sources detected within the 15.6 deg$^2$ footprint of the {\it XMM-Newton} AO13 Stripe 82 survey area; we also remove the  archival X-ray sources from the eBOSS program that do not overlap the AO13 survey area. Table \ref{dem_ao13fov} provides a demographic summary of the X-ray and {\it WISE} sources used in this analysis, amounting to 2751 AGN total.

\begin{deluxetable*}{lrrrr}
  \tablewidth{0pt}
  \tablecaption{\label{dem_ao13fov} Classification Summary of Sources that Overlap the 15.6 deg$^2$ {\it XMM-Newton} AO13 Footprint\tablenotemark{1}}
  \tablehead{\colhead{Classification} & \colhead{X-ray AGN} & \colhead{{\it WISE} AGN} & \colhead{X-ray \& {\it WISE} AGN} & \colhead{X-ray or {\it WISE} AGN}}

  Type 1 AGN             &  1413     & 1172 & 639 & 1946 \\
  Optically Obscured AGN &  361     &  495 &  52  &  804
  \enddata
  \tablenotetext{1}{There is an additional X-ray AGN that lacks an optical spectroscopic classification in the archival catalogs.}
\end{deluxetable*}

We immediately see from Table \ref{dem_ao13fov} that the space density is comparable between the {\it WISE} AGN ($\sim$108/deg$^2$) and the X-ray AGN ($\sim$114/deg$^2$). Only $\sim$23\% of the X-ray or {\it WISE} AGN in this sample are classified as AGN on the basis of optical broad lines, X-ray luminosity, and MIR colors (639 out of 2751). Of the 804 obscured AGN, only $\sim$6\% are identified as accreting black holes on the basis of both X-ray emission and red MIR colors.

\begin{figure}[ht]
  \begin{center}
  \includegraphics[scale=0.43]{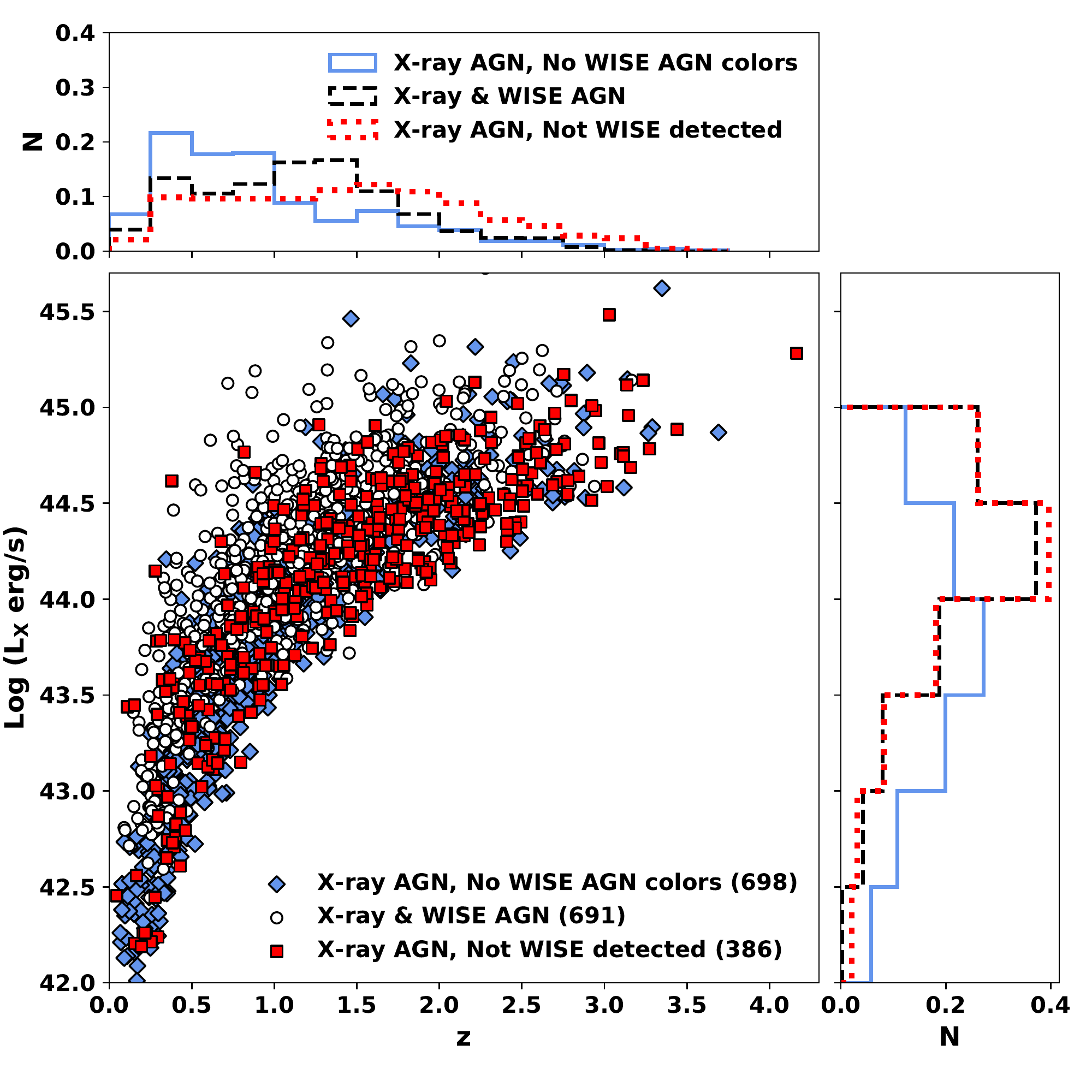}~
  \caption{\label{comp_xagn_w_wise} X-ray luminosity as a function of redshift for X-ray AGN that are detected by {\it WISE} and have $W1-W2$ colors bluer than the \citet{assef2018} 75\% reliability color cut (blue diamonds; solid blue histogram), have $W1 - W2$ colors that obey the \citet{assef2018} color criterion (white circles; black dashed histogram), and those undetected by {\it WISE} (red squares; red dotted histogram). The X-ray AGN with blue {\it WISE} colors tend to have moderate X-ray luminosities (i.e., $10^{42.5} < L_{\rm X} < 10^{44}$ erg s$^{-1}$) and lie below $z < 1$, while those undetected by {\it WISE} populate the same parameter space as those with {\it WISE} AGN colors: a not insignificant fraction of luminous black hole growth is missed by $W1-W2$ selection due to {\it WISE} non-detections. Normalized histograms are shown to better illustrate parameter space spanned by these source populations.}
  \end{center}
\end{figure}

\begin{figure}[ht]
  \begin{center}
  \includegraphics[scale=0.43]{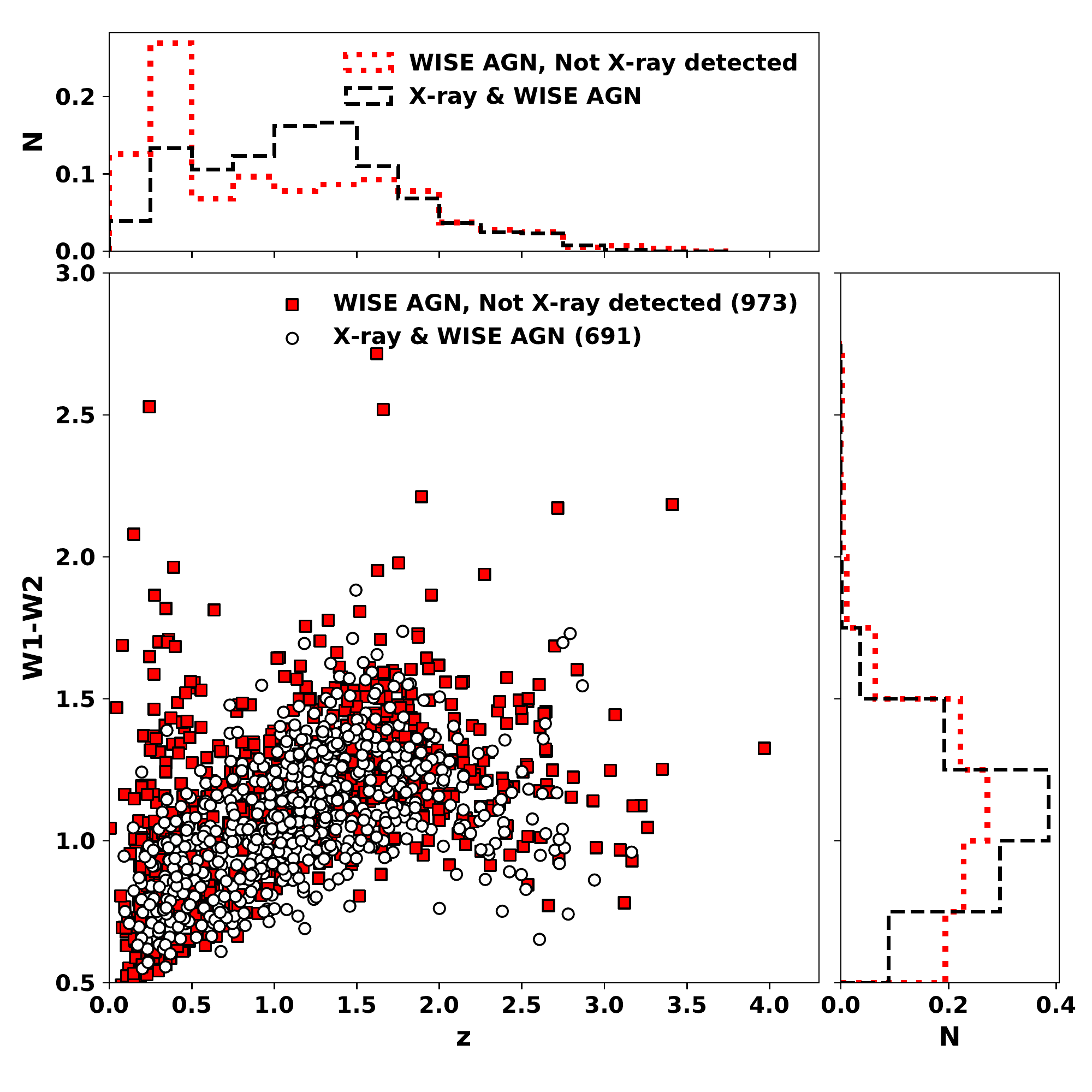}
  \caption{\label{comp_wagn_w_xray} MIR $W1-W2$ color as a function of redshift for the {\it WISE} AGN, with those sources also identified as AGN by their X-ray luminosity (white circles; black dashed histogram), and those completely undetected in X-rays (red squares; red dotted histogram), shown; not shown are the 3 X-ray sources with X-ray luminosities below 10$^{42}$ erg s$^{-1}$. 58\% of the {\it WISE} AGN are not detected in X-rays. No $W1-W2$ color trend is seen between X-ray detections and non-detections, as would be expected if the {\it WISE} AGN not detected in X-rays were more obscured. Normalized histograms are shown to better illustrate parameter space spanned by these source populations.}
  \end{center}
\end{figure}

In Figures \ref{comp_xagn_w_wise} and \ref{comp_wagn_w_xray}, we explore the populations detected and missed by defining AGN based on X-ray luminosity and MIR color, highlighting the complementarity of multiple selection criteria to offer a comprehensive view of black hole growth. 

The 1775 X-ray AGN are shown in Figure \ref{comp_xagn_w_wise}, where we highlight the subset of sources also identified as AGN on the basis of their $W1-W2$ colors (691; 39\%), those AGN detected by {\it WISE} but with bluer $W1-W2$ colors than the \citet{assef2018} 75\% reliability color cut (698; 39\%), and those undetected by {\it WISE} (386; 22\%). Similar to trends previously reported in other samples \citep[e.g.,][]{eckart,mendez,menzel,lamassa2016b}, the X-ray AGN detected by {\it WISE} that do not meet the $W1-W2$ color criterion tend to be at low to moderate X-ray luminosities (i.e., 64\% have $L_{\rm X} < 10^{44}$ erg s$^{-1}$). This percentage is consistent with the 50-70\% of X-ray AGN not identified as such by their $W1-W2$ colors found by \citet{georgakakis}, albeit with a more conservative color cut ($W1-W2 > 0.8$) than what we use here. It is reasonable to assume that in these cases, the AGN is not dominating the MIR emission, which is a population to which the $W1-W2$ color selection is not tuned.

However, we also find that though MIR-selected AGN are found at the highest X-ray luminosities and redshifts, the X-ray sources {\it undetected} by {\it WISE} populate the same parameter space. About a third of the highest luminosity ($L_{\rm X} > 10^{44}$ erg s$^{-1}$), highest redshift ($z > 1$) X-ray AGN are undetected by {\it WISE}, indicating that MIR selection can miss a non-negligible fraction of the most luminous black hole growth. Many of these sources may be ``hot dust poor quasars'' described in \citet{hao2010,hao2011} and \citet{lyu}. This population was shown to have anomalously weak rest-frame NIR emission between 1-3$\mu$m compared to other Type 1 AGN, which can be explained by a low dust covering factor. Indeed, the rest-frame $W1$ and $W2$ passbands at $z>1$ probe rest-frame NIR emission, suggesting this interpretation has some merit. However, fitting the broad-band SEDs of these sources to derive NIR and optical slopes are required to test whether they fit the definition of hot dust poor quasars, which will be explored in a follow-up paper.

We perform the corollary analysis in Figure \ref{comp_wagn_w_xray}, where we investigate the $W1-W2$ color as a function of redshift for the 691 {\it WISE} and X-ray AGN and the 973 {\it WISE} AGN not detected in X-rays (only 3 sources classified as {\it WISE} AGN are X-ray galaxies), representing 42\% and  58\% of the {\it WISE} AGN, respectively. Though it seems plausible that the {\it WISE} AGN undetected by X-rays suffer from high levels of extinction, we do not observe that these sources have redder $W1-W2$ colors compared with the X-ray detected AGN.

\section{Conclusions}
We reported on the results of an SDSS-IV eBOSS spectroscopic survey that covered 36.8 deg$^2$ of Stripe 82 in the fall of 2015. About half of this survey area (15.6 deg$^2$) overlaps the largest, contiguous region of the Stripe 82 X-ray survey, observed by {\it XMM-Newton} in AO13 \citep[PI: Urry;][]{lamassa2016a}. The primary targets of this survey were X-ray sources and {\it WISE} AGN candidates from the ALLWISE survey \citep{Mainzer2011} identified on the basis of their $W1-W2$ colors \citep[i.e., the 75\% reliabibility threshold of][]{assef2013}. SDSS counterparts to the X-ray and {\it WISE} sources were identified using the statistical maximum likelihood approach, as described in \citep{lamassa2016a} for the X-ray sources and in the main text for the {\it WISE} population. Additional ``filler'' targets were observed to make use of all available spectroscopic fibers across the SDSS plates. Subsequent to the SDSS-IV eBOSS observing program, an updated multi-wavelength matched catalog to the Stripe 82X survey was published by \citet{ananna}, and the $W1-W2$ color criteria for AGN were modified with respect to the eBOSS target list we created \citep{assef2018}, leading us to curate the sources in our final published catalog with respect to the objects we targeted.

In total, 2262 SDSS counterparts to the original X-ray and {\it WISE} AGN candidate target lists were spectroscopically observed. We visually inspected all spectra, finding that 1769 sources (78\%) were of sufficient quality to determine redshifts and classifications,  where we achieved a $\sim$37\% identification rate at the faintest magnitudes  ($22.5 < r < 23.5$; Figure \ref{success_rmag}). We recommend visual inspection of spectra flagged by \textsc{zwarning} or having S/N $<$ 2.25 to maximize sample size and reliability of results; only 3\% of sources not flagged by \textsc{zwarning} and with S/N $>$ 2.25 were found to have discrepant redshifts and/or classifications between visual inspection and the pipeline. If limited resources preclude visual inspection of spectra, then imposing a S/N threshold exceeding 2.25 and a null \textsc{zwarning} flag would result in a reliable sample, but at the expense of discarding $\sim$30\% of spectra that would otherwise be of sufficient quality for analyzing source demographics.

After vetting the SDSS spectroscopic results and curating the target lists to only include X-ray/SDSS counterparts from \citet{ananna} and {\it WISE} AGN meeting the \citet{assef2018} $W1-W2$ color criteria at the 75\% level, we combined this sample with X-ray and {\it WISE} AGN in the survey area with spectroscopic redshifts from other SDSS programs \citep{sdssdr7,sdssdr8,alam2015, albareti, sdssdr14, paris, dr14q}, 2SLAQ \citep{croom}, 6dF \citep{jones2004, jones2009}, and dedicated follow-up programs of Stripe 82 X-ray sources \citep{lamassa2016a, lamassa2017}. The total sample is 82\% complete to $r \sim 22$, with the X-ray and {\it WISE} AGN samples being $\sim$88\% and $\sim$82\% complete at this magnitude limit (Figure \ref{spec_compl}).

Our spectroscopic sample consists of 4847 sources, of which 4730 are AGN (1790 X-ray AGN, 3638 {\it WISE} AGN, 698 X-ray and {\it WISE} AGN): 3310 are Type 1 AGN (70\%) and 1418 are optically obscured AGN (30\%; Table \ref{dem_summary}); two AGN did not have spectroscopic classifications in the archival catalogs we queried. A vast majority of the optically obscured AGN (76\%) were identified via the eBOSS Stripe 82X survey. The AGN range in redshift from $0.02 < z < 4.2$, with a median redshift of $z \sim 1$ (Figure \ref{z_distr}). BPT analysis of the $z < 0.5$ AGN with high S/N emission lines show that 50\% of the {\it WISE} AGN occupy the star-forming locus (Figure \ref{bpt_all}): whether these sources are optically buried AGN or star-forming galaxies whose MIR colors mimic those of AGN requires further analysis \citep[e.g.][]{satyapal}.

In the 15.6 deg$^2$ area of the {\it XMM-Newton} AO13 Stripe 82 footprint, we  compared the AGN populations from X-ray and MIR-selection (Table \ref{dem_ao13fov}), finding the following trends among the 2751 AGN in this area-restricted sample:
\begin{itemize}
\item only $\sim$6\% of the optically obscured AGN (52 out of 804) are both X-ray and {\it WISE} AGN, highlighting the importance of both X-ray and MIR selection to recover AGN missed by optical surveys;
\item 61\% of X-ray AGN (1084 out of 1775) are not MIR AGN (Figure \ref{comp_xagn_w_wise}):
  \begin{itemize}
  \item 39\% are detected by {\it WISE} but have $W1-W2$ colors too blue for the \citet{assef2018} 75\% reliability color definition. These sources are generally at lower luminosity, where the AGN is not contributing signficantly to the MIR SED \citep[see also, e.g.,][]{eckart,mendez,menzel,lamassa2016b}.
  \item 22\% are {\it undetected} by {\it WISE}. These sources are generally X-ray luminous (i.e., $L_{\rm x} > 10^{44}$ erg s$^{-1}$), challenging the conventional wisdom that MIR color selection identifies all luminous AGN: these sources may have anomalous dust properties, similar to ``hot dust poor'' or ``hot dust deficient'' quasars \citep{hao2010,hao2011,lyu}.
  \end{itemize}
 \item 58\% of {\it WISE} AGN (973 out of 1664) are not detected in X-rays (Figure \ref{comp_wagn_w_xray}), but there is no clear $W1-W2$ color difference between X-ray AGN and non-detections, indicating that the {\it WISE} AGN undetected in X-rays are not redder and hence may not preferentially be more obscured.
  \end{itemize}

\acknowledgements{We thank the referee for a careful reading of the manuscript and for providing insightful comments that helped us streamline the paper.

  Funding for the Sloan Digital Sky Survey IV has been provided by the Alfred P. Sloan Foundation, the U.S. Department of Energy Office of Science, and the Participating Institutions. SDSS-IV acknowledges
support and resources from the Center for High-Performance Computing at
the University of Utah. The SDSS web site is www.sdss.org.

SDSS-IV is managed by the Astrophysical Research Consortium for the 
Participating Institutions of the SDSS Collaboration including the 
Brazilian Participation Group, the Carnegie Institution for Science, 
Carnegie Mellon University, the Chilean Participation Group, the French Participation Group, Harvard-Smithsonian Center for Astrophysics, 
Instituto de Astrof\'isica de Canarias, The Johns Hopkins University, 
Kavli Institute for the Physics and Mathematics of the Universe (IPMU) / 
University of Tokyo, Lawrence Berkeley National Laboratory, 
Leibniz Institut f\"ur Astrophysik Potsdam (AIP),  
Max-Planck-Institut f\"ur Astronomie (MPIA Heidelberg), 
Max-Planck-Institut f\"ur Astrophysik (MPA Garching), 
Max-Planck-Institut f\"ur Extraterrestrische Physik (MPE), 
National Astronomical Observatories of China, New Mexico State University, 
New York University, University of Notre Dame, 
Observat\'ario Nacional / MCTI, The Ohio State University, 
Pennsylvania State University, Shanghai Astronomical Observatory, 
United Kingdom Participation Group,
Universidad Nacional Aut\'onoma de M\'exico, University of Arizona, 
University of Colorado Boulder, University of Oxford, University of Portsmouth, 
University of Utah, University of Virginia, University of Washington, University of Wisconsin, 
Vanderbilt University, and Yale University.

This research made use of Astropy, a community-developed core Python package for Astronomy \citep{astropy}.

Some of the data presented herein were obtained at the W. M. Keck Observatory, which is operated as a scientific partnership among the California Institute of Technology, the University of California and the National Aeronautics and Space Administration. The Observatory was made possible by the generous financial support of the W. M. Keck Foundation. The authors wish to recognize and acknowledge the very significant cultural role and reverence that the summit of Maunakea has always had within the indigenous Hawaiian community.  We are most fortunate to have the opportunity to conduct observations from this mountain.

SML acknowledges funding support from 17-ADAP17-0055.}

\facility{Sloan, XMM, CXO, Hale, WIYN, Keck:I (LRIS), Gemini:Gillet}

\appendix

\section{Comparison between Stripe 82X AGN and  XMM-XXL-N AGN}

We compare the X-ray AGN identified in this program with those observed in a similar SDSS BOSS program to target X-ray AGN from the 18 deg$^2$ XMM-XXL northern field \citep{menzel}. With a flux limit of $\sim 3.5\times 10^{-15}$ erg s$^{-1}$ cm$^{-2}$ at 50\% of the survey area, it is slightly more sensitive than the {\it XMM}-Newton AO13 component of the Stripe 82X survey ($F_{\rm 0.5-2keV,lim} \sim 5 \times 10^{-15}$ erg s$^{-1}$ cm$^{-2}$ at half the survey area). Out of 8445 X-ray point sources in the XMM-XXL-N field \citep{pierre2016,liu2016}, 3004 were spectroscopically followed up by BOSS, garnering reliable redshifts and spectroscopic classifications for 2514 extragalactic sources (including two blazars) and 85 stars. We {\it a priori} remove the 57 sources with ``not classifiable'' BOSS spectra from the \citet{menzel} catalog in our comparison below, since though their redshifts are securely determined via visual inspection, the low S/N of the spectra precludes optical spectroscopic classification.

We follow our categorization scheme above when classifying the sources in \citet{menzel}, where the optical spectroscopic classification of ``Type 1 AGN'' refer to objects that have at least one broad emission line and ``Optically Obscured AGN'' indicate sources with narrow emission lines only or absorption lines, regardless of their BPT designation. We consider these objects X-ray AGN if their estimated $k$-corrected 2 - 10 keV luminosity ($L_{\rm X}$) exceeds 10$^{42}$ erg s$^{-1}$, where we follow a similar prescription to that of the Stripe 82X sources to calculate $L_{\rm X}$ if the source is not detected in hard band. Here, we use $\Gamma=1.4$ to calculate $k$-corrected luminosities since this was the spectral slope assumed by \citet{liu2016} when converting from counts to fluxes. Based on this spectral slope, we use a correction factor of 0.74 and 2.88 to convert the luminosity from the 0.5 - 10 keV and 0.5 - 2 keV bands, respectively, to the 2 - 10 keV band.

In XMM-XXL-N, we find 1787 Type 1 AGN and 654 optically obscured AGN. The detection threshold for the XMM-XXL-N sources in \citet{liu2016} is lower than that used in Stripe 82X, with a detection significance set to $P < 4 \times 10^{-6}$ for XMM-XXL-N, compared with $P < 3 \times 10^{-7}$ for Stripe 82X. Due to the lower significance threshold of XMM-XXL-N and the deeper observations, the number density of spectroscopically confirmed X-ray AGN in the 18 deg$^2$ XMM-XXL-N ($\sim$136 deg$^{-2}$) is higher than that of the 15.6 deg$^2$ Stripe 82X {\it XMM}-AO13 field ($\sim$114 deg$^2$).

\begin{figure*}[ht]
  \begin{center}
  \includegraphics[scale=0.55]{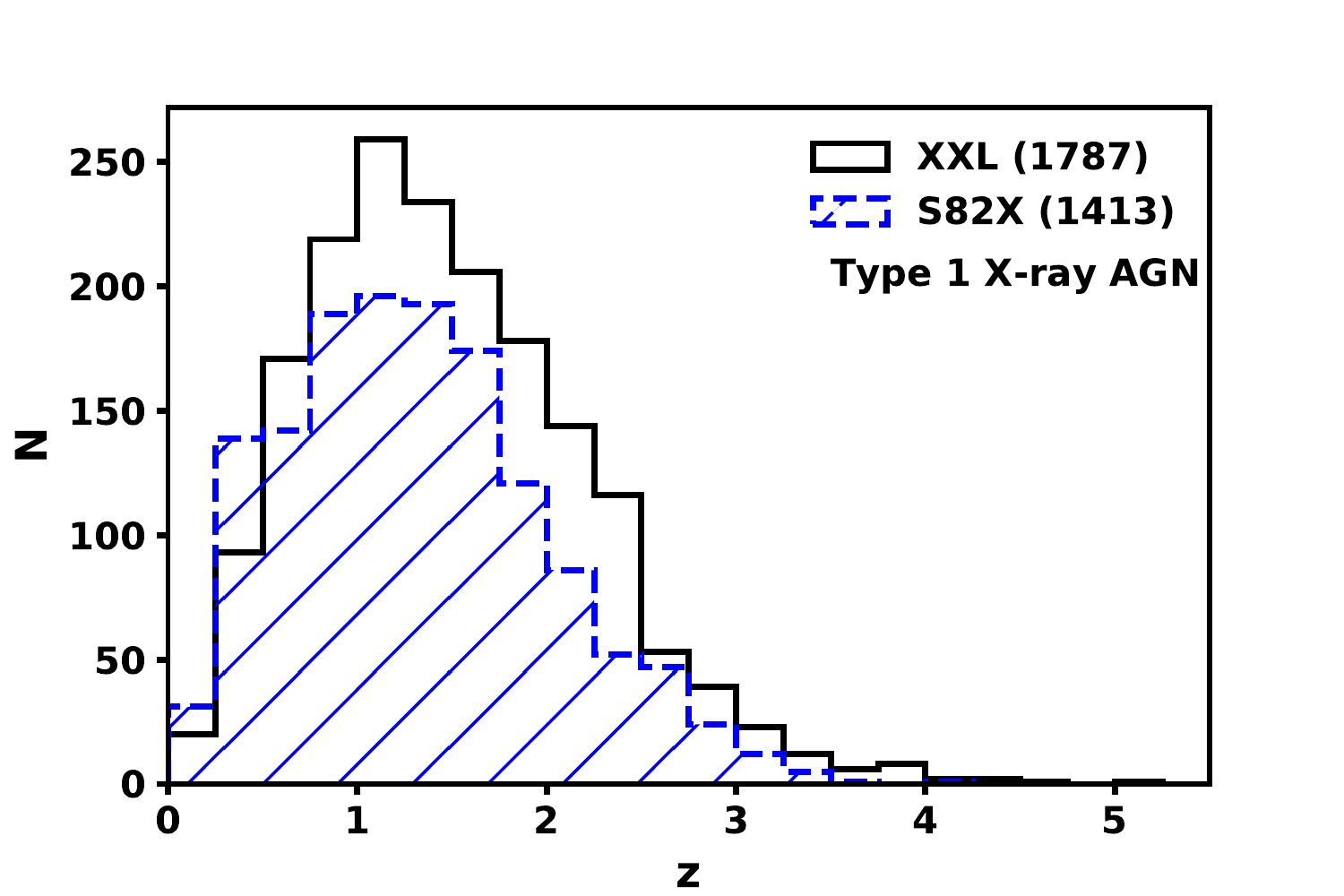}~
  \includegraphics[scale=0.55]{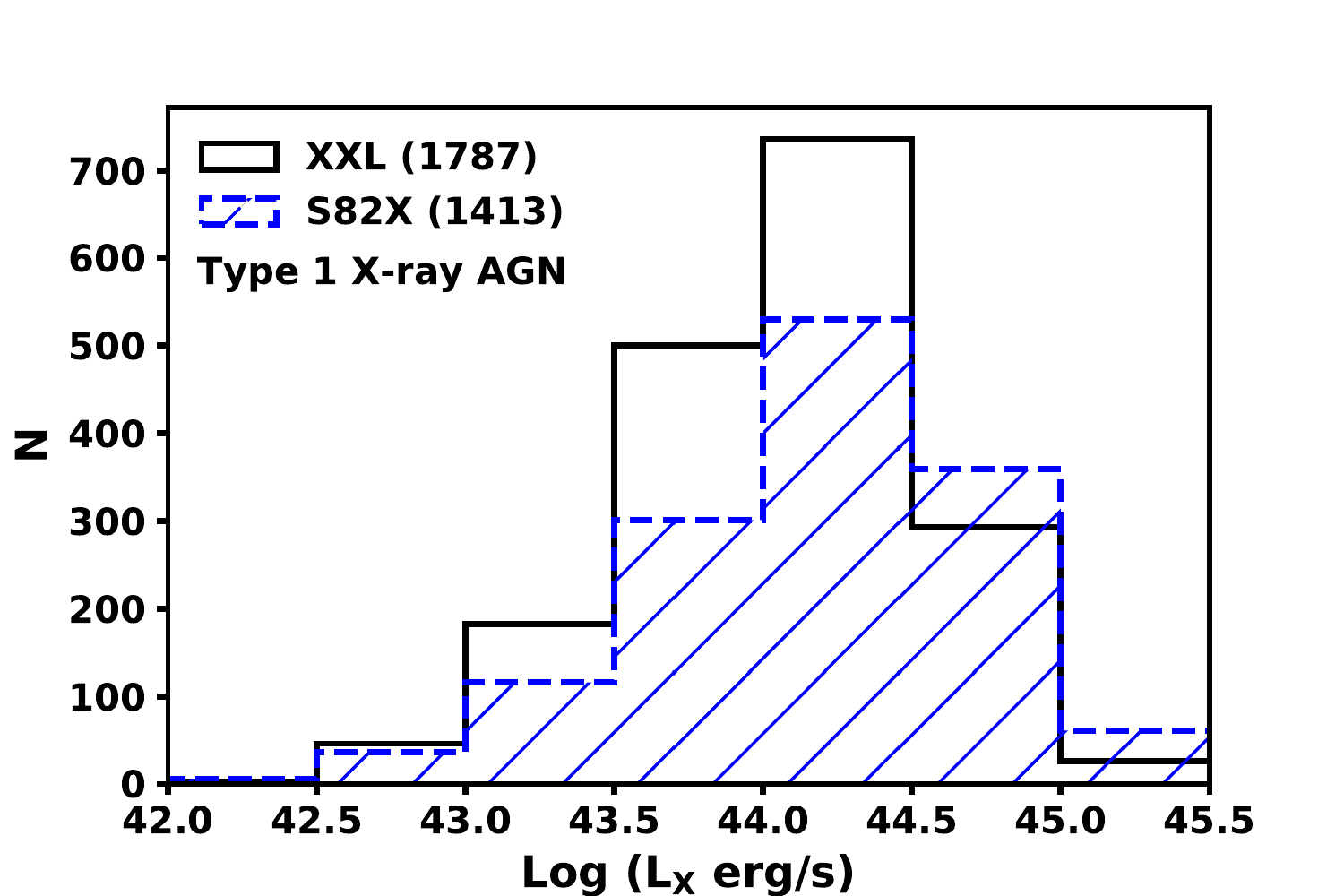}
  \includegraphics[scale=0.55]{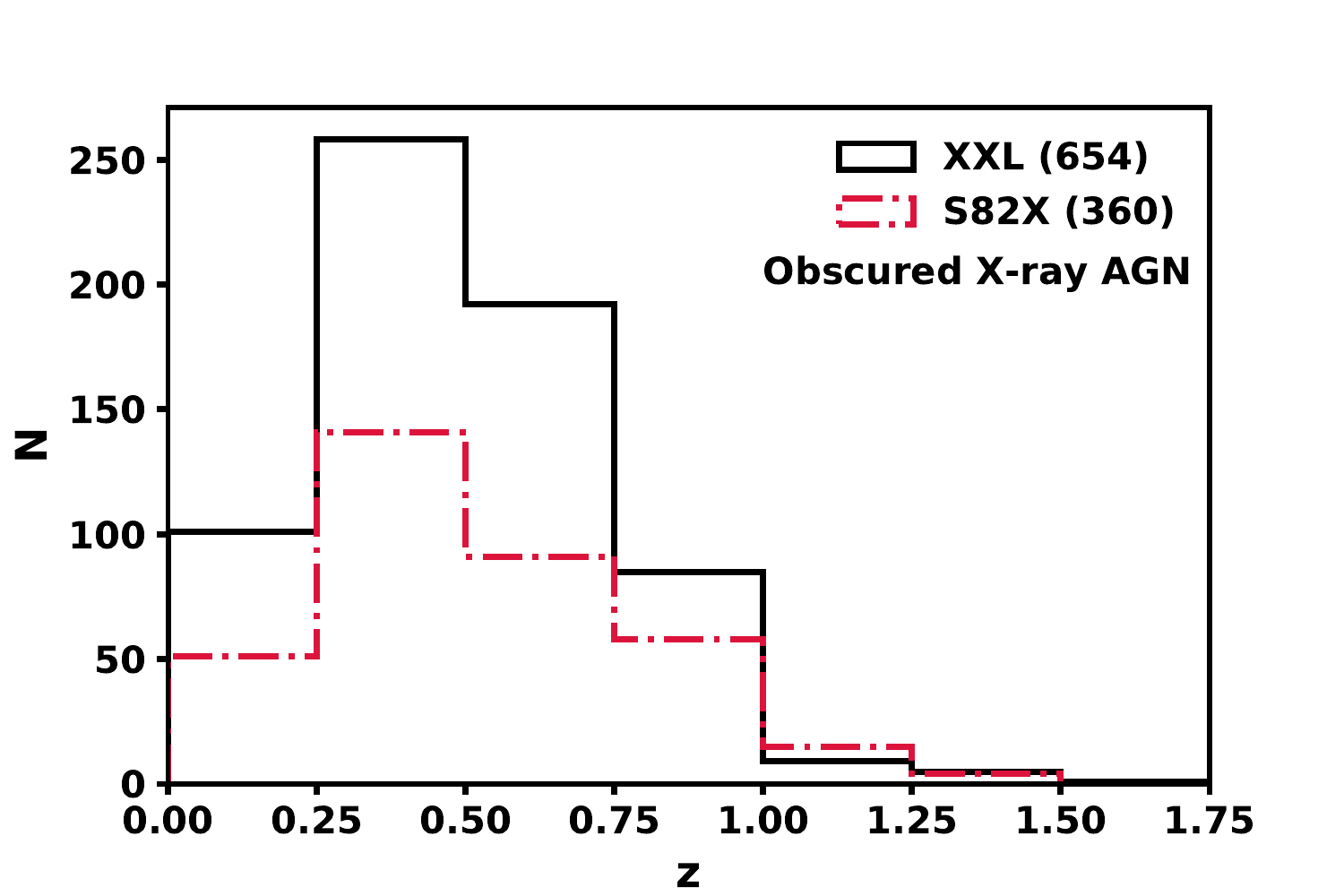}~
  \includegraphics[scale=0.55]{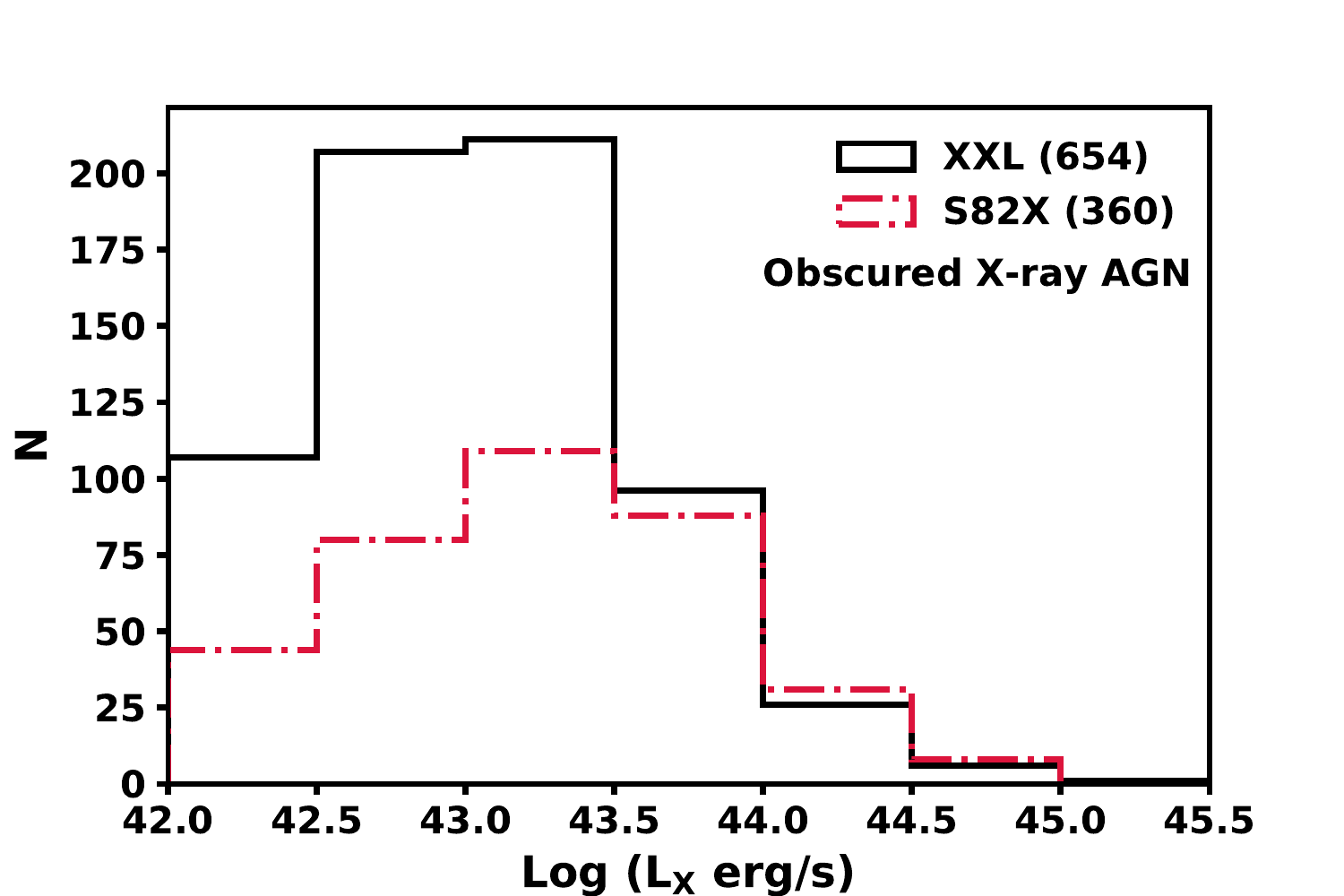}
  \caption{\label{xxl_v_s82x} ({\it Left}) Redshift and ({\it right}) X-ray luminosity  distribution for ({\it top}) Type 1 and ({\it bottom}) optically obscured X-ray AGN from the 18 deg$^2$ XMM-XXL-N survey \citep[black, solid line;][]{pierre2016,liu2016,menzel} and the 15.6 deg$^2$ {\it XMM-Newton} AO13 portion of the Stripe 82X survey (blue hatched histogram for Type 1 AGN; red dot-dashed histogram for optically obscured AGN). Though XMM-XXL-N has a higher number density of AGN due to deeper coverage and a lower detection threshold, the number of AGN detected at the highest redshifts and luminosities are comparable between the two surveys, demonstrating the necessity of wide areal coverage to build statistics in this parameter space.}
  \end{center}
\end{figure*}

We find a smaller fraction of optically obscured X-ray AGN between Stripe 82X (20\%) compared with XMM-XXL-N (27\%). However, despite the higher number density of AGN in XMM-XXL-N, the number of high-redshift ($z > 2.5$ for Type 1 AGN; $z > 0.75$ for optically obscured AGN) and high-luminosity ($L_{\rm X}> 10^{44.5}$ erg s$^{-1}$ for Type 1 AGN; $L_{\rm X} > 10^{43.5}$ erg s$^{-1}$ for optically obscured AGN) are similar, as demonstrated in Figure \ref{xxl_v_s82x}. Hence, wide area at moderate depths ($\sim 5 \times 10^{-15}$ erg s$^{-1}$ cm$^{-2}$) is more important than deeper coverage in unveiling AGN at the highest luminosities and redshifts.

\end{document}